\documentclass[journal]{IEEEtran}

\ifCLASSINFOpdf
\else
\fi

\usepackage{booktabs}
\usepackage{xcolor}
\usepackage{multirow}
\usepackage{tikz}
\usetikzlibrary{arrows.meta,positioning,calc,decorations.pathreplacing,shapes.geometric,fit}
\usepackage{subcaption}
\usepackage{array, tabularx}
\usepackage{pifont}
\usepackage{amsmath,amssymb}
\usepackage{makecell}
\usepackage{amssymb}
\usepackage{amsmath}
\usepackage{booktabs}
\usepackage{multirow}
\usepackage{xcolor}
\usepackage{graphicx}
\usepackage{url}
\usepackage{pifont}
\usepackage{graphicx}
\usepackage{float}
\usepackage[markup=underlined]{changes}

\newcommand{\cmark}{\textcolor{black}{\scalebox{1.2}{$\checkmark$}}}
\newcommand{\xmark}{\textcolor{black}{\scalebox{1.2}{$\times$}}}

\definechangesauthor[name={Arman}, color=blue]{AD}

\begin{document}

\title{METIS: A Declarative Slice Orchestrator for Application-Centric 5G/6G Networks}

\author{\IEEEauthorblockN{Arman Divband\IEEEauthorrefmark{1},
Ali Yaghoubian\IEEEauthorrefmark{1}, and
Navid Nikaein\IEEEauthorrefmark{1}}
\IEEEauthorblockA{\IEEEauthorrefmark{1}\textit{EURECOM}, Sophia-Antipolis, France\\
Email: \{Arman.Divband, Ali.Yaghoubian, Navid.Nikaein\}@eurecom.fr}
\thanks{Corresponding authors: Arman Divband (email: Arman.Divband@eurecom.fr))}}

\maketitle

% \IEEEtitleabstractindextext{%
\begin{abstract}
Network slicing is the cornerstone of application-aware 5G-Advanced and 6G networks, yet dynamic lifecycle management of network slice instances with coordinated quality-of-service enforcement across the radio access network and core network remains an unresolved challenge. Existing orchestrators rely on network-centric data models, imperative workflows, and static slice templates, while O-RAN addresses radio-side slice control independently of 3GPP core-side control, leaving slice-level quality-of-service enforcement uncoordinated across domains. This paper introduces METIS, a declarative slice orchestrator that manages the complete Day-0/1/2 lifecycle of network slice instances through cascaded reconciliation loops. METIS defines an application-centric data model for service profiles, enabling customers to describe the semantics and quality-of-experience requirements of their applications. From these, METIS derives 3GPP-aligned slice profiles via hierarchical aggregation following the 5G quality-of-service model, eliminating static templates, and jointly coordinates O-RAN and 3GPP slicing for slice instantiation and enforcement. Our central finding is a structural asymmetry in end-to-end slice control: downlink traffic can be shaped at the core before reaching the radio access network, but uplink leaves the user equipment unregulated, so core-only slicing cannot reliably satisfy uplink service-level agreements---radio-side enforcement is necessary, not merely complementary. Evaluated on a 5G cloud-native testbed in a campus-event scenario, METIS completes slice creation, update, upgrade, and deletion within 22.4, 5.1, 52.2, and 32.1 seconds, respectively; sustains full service-level-agreement satisfaction under concurrent multi-slice overload; scales to 63 slice instances across nine zones consuming under 0.03 processor cores total; and recovers slices from injected failures across four levels in under 19 seconds.
\end{abstract}


% }

\section{Introduction} \label{sec:introduction}

As envisioned in Recommendation ITU-R M.2160-0~\cite{ITU-R-M2160}, 5G-Advanced and 6G networks are expected to support services with diverse requirements in terms of latency, reliability, throughput, security, coverage, and energy efficiency. Network slicing is a primary enabler for creating customized and isolated logical networks, Network
Slice Instances (NSIs), over a shared physical infrastructure, allowing each service to receive the specific resources and performance guarantees it requires. However, supporting a wide range of services with varying lifespans makes dynamic NSI lifecycle management a key operational challenge. This challenge stems from the reliance on network-centric data models, which typically leads to static, template-based slice descriptors and limits the ability of Communication
Service Customers (CSCs)~\cite{ts28530} to dynamically express evolving service requirements throughout the NSI lifecycle without requiring expertise in 5G-Advanced or 6G technologies.

To automate NSI lifecycle management, 3GPP TS 28.533~\cite{ts28533} defines an architecture that integrates the Network Slice Management Function (NSMF)~\cite{ts28530} with the ETSI Network Functions Virtualisation (NFV) Management and Orchestration (MANO) framework~\cite{etsi_gs_nfv_006}. In this architecture, CSCs express the communication service requirements---formulated largely as 5G network slice attributes---to the Communication Service Management Function (CSMF)~\cite{ts28530}, which captures them as Service Profiles for the NSMF to derive the corresponding Slice Profiles and then orchestrate the associated NSIs. The O-RAN specification~\cite{oran_slicing_arch} reuses these 3GPP-defined management functions within its Service Management and Orchestration (SMO) framework~\cite{oran_oam_arch} to support NSI orchestration in the Radio Access Network (RAN) domain.

Although 3GPP TS 28.541~\cite{ts28541} specifies the data model for Service Profiles, this model is network-centric (expressing 5G network-slice service-requirement attributes) rather than application-centric (capturing the semantic and Quality of Experience (QoE) requirements of the application). Moreover, while O-RAN reuses the 3GPP management functions, its RAN slicing control operates independently of 3GPP-defined Core Network (CN) slicing: the Policy Control Function (PCF) governs Quality of Service (QoS) policies in the CN while the Non-Real-Time (Non-RT) and Near-Real-Time (Near-RT) RAN Intelligent Controllers (RICs) do so in the RAN, with no coordination mechanism between the two domains---yet aligning them is crucial to establish and preserve slice-level QoS throughout the NSI lifecycle.

Existing NSI orchestration frameworks, such as the Open Network Automation Platform (ONAP), rely on imperative, workflow-driven principles, characterized by step-by-step deployment procedures and intricate state machines~\cite{onap_so_architecture,onap_so_github}. This design style is also reflected in the operation- and procedure-oriented provisioning model of 3GPP TS 28.531~\cite{ts28531}, which specifies NSI lifecycle actions through allocation, modification, activation, deactivation, and deallocation procedures. Such approaches remain insufficient for enabling a zero-touch NSI orchestration framework that dynamically performs NSI lifecycle management operations in an idempotent fashion, i.e., re-applying the same operation produces the same NSI state as applying it once, without side effects on the NSI or on other NSIs sharing its Network Functions (NFs). Specifically, if a single step in a deployment procedure fails, the orchestration process may stall and require manual intervention~\cite{azure_compensating_transaction}. Although recent cloud-native frameworks, Nephio~\cite{nephio} and Athena~\cite{athena}, adopt declarative and reconciliation-driven principles, their orchestration scope remains primarily limited to NFs and Network Services (NSs), respectively, rather than treating Service Profiles, Slice Profiles, and NSIs as lifecycle-managed resources.

To bridge these gaps, this paper presents METIS, a cloud-native NSI orchestration framework that treats NSIs as declarative, application-driven resources spanning the RAN and CN domains. METIS introduces a new, application-centric data model for Service Profiles that captures the semantic and QoE requirements of services directly from CSCs, from which 3GPP-aligned Slice Profiles are automatically derived. Its key design principle is to distribute NSI orchestration logic across cascaded reconciliation loops, where each loop manages one abstraction level and propagates its desired state to the loops below. These cascaded loops coordinate O-RAN and 3GPP slicing providers for domain-specific NSI orchestration across a geographical area and enforce slice-level QoS through a coordinated control plane. They allow METIS to dynamically perform Day-0/1/2 NSI lifecycle operations, reconciling toward the desired state rather than stalling under failures. 

\indent In summary, this paper makes the following contributions:
\begin{itemize}
    \item \textbf{Application-centric Service Profiles:} A data model that captures the semantic information and QoE requirements of applications by focusing on application-level attributes rather
    than network-centric slice attributes, and automatically derives QoS profiles of 3GPP-aligned Slice Profiles following the 5G QoS model (Subsection~\ref{subsec:day0}).

    \item \textbf{Joint O-RAN and 3GPP slicing:} A declarative design coordinating O-RAN and 3GPP slicing providers for NSI orchestration and QoS enforcement across the RAN and CN domains (Sections~\ref{Sec:design_overview} and~\ref{Sec:systemdesign_and_implementatoin}).
    
    \item \textbf{Full Day-0/1/2 NSI lifecycle management:} A novel multi-level operator design based on cascaded reconciliation loops for dynamic NSI creation, termination, update, and upgrade while preserving idempotency (Section~\ref{Sec:systemdesign_and_implementatoin}).

    \item \textbf{Prototype and experimental validation of METIS:} A concrete implementation validated through a realistic use case (campus event) covering dynamic NSI lifecycle management and joint RAN--CN slicing with QoS enforcement. The evaluation uses a set of well-defined NSI lifecycle metrics and considers both scalability and failure-recovery experiments (Subsection~\ref{sec:implementation} and Section~\ref{Sec:evaluation}).
  
\end{itemize}

\section{Related Work} \label{Sec:related_works}

\textbf{Management and Orchestration.}
Early MANO frameworks such as OSM~\cite{osm} and OpenStack Tacker~\cite{tacker} were designed in line with the ETSI NFV MANO architecture~\cite{etsi_nfv_man001} and primarily focused on orchestrating Virtualized Network Functions (VNFs) to support the lifecycle management of NSs. With the emergence of cloud-native technologies, the ETSI NFV MANO specifications~\cite{etsi_nfv_ifa036, etsi_nfv_ifa040} were extended to support the orchestration of CNFs. Consequently, OSM and OpenStack Tacker extended their existing imperative and workflow-driven orchestration logic~\cite{osm_source_code,tacker_source_code} to operate over Kubernetes, rather than redesigning it around declarative and reconciliation-driven principles. 

%%%%%%%%%%%%%%%%%%%%%%%%%%%%%
%%%%%%%%%%%%%%%%%%%%%%%%%%%%%
\begin{table}[!t]
\centering
\caption{Classification of Network Slice Orchestration Frameworks}
\label{tab:orchestrator-classification}
\renewcommand{\arraystretch}{1.1}
\setlength{\tabcolsep}{2.5pt}
\scriptsize
\resizebox{\columnwidth}{!}{%
\begin{tabular}{|c|c|c|c|c|}
\hline
\textbf{Solution} &
\shortstack{\textbf{NSI} \\ \textbf{Orch.}} &
\shortstack{\textbf{App-Centric} \\ \textbf{Service Profile}} &
\shortstack{\textbf{Control Logic}} &
\textbf{LCM} \\
\hline
OSM\cite{osm}    & \xmark     & \xmark & IW & D-0/1 \\
\hline
Tacker\cite{tacker} & \xmark     & \xmark & IW & D-0/1 \\
\hline
Nephio\cite{nephio} & \xmark     & \xmark & DR & D-0/1 \\
\hline
Athena\cite{athena} & \xmark     & \xmark & DR & D-0/1 \\
\hline
MANO–OSS/BSS\cite{free5gmano} & Partial/\checkmark & \xmark & IW & D-0/1/pD-2 \\
\hline
ORANSlice\cite{oranslice} & Partial/\checkmark & \xmark & IW & pD-2 \\
\hline
CLiSO\cite{cliso}  & \checkmark & \xmark & IW & D-0/1/pD-2 \\
\hline
NASP\cite{nasp}   & \checkmark & \xmark & IW & D-0/1/pD-2 \\
\hline
ONAP\cite{onap_so_architecture}   & \checkmark & \xmark & IW & D-0/1/pD-2 \\
\hline
\textbf{METIS} &    \textbf{\checkmark} &   \textbf{\checkmark} &   \textbf{DR} &   \textbf{D-0/1/2} \\
\hline
\end{tabular}%
}
\footnotesize{IW: Imperative Workflow-driven logic; DR: Declarative Reconciliation-driven logic; pD-2: partial Day-2 support.}
\vspace{-1em}
\end{table}
%%%%%%%%%%%%%%%%%%%%%%%%%
%%%%%%%%%%%%%%%%%%%%%%%%%

To bridge this gap, recent orchestration frameworks such as Nephio~\cite{nephio} and Athena~\cite{athena} employ orchestration logic based on declarative and reconciliation-driven principles. Nephio orchestrates CNFs across multi-cluster and multi-vendor environments, while Athena performs the lifecycle management of NSs by orchestrating O-RAN and CN CNFs using two hierarchical Operators and a unified data model that serves as a descriptor for CNFs. Despite representing a new generation of MANO frameworks, both Nephio~\cite{nephio_source_code} and Athena remain focused on CNF and NS orchestration. Although Athena states that it "natively accepts slices as first-class citizens and performs full lifecycle operations on them"~\cite{athena}, its treatment of slices remains at a different level of abstraction from NSI lifecycle management: it does not explicitly address Service Profiles, Slice Profiles, or NSIs as lifecycle-managed entities, nor does it provide an evaluation focused on NSI lifecycle management. 

\textbf{Slice Orchestration.}
Existing slice orchestration works can be broadly categorized according to whether they address slice management within a single network domain or provide end-to-end NSI lifecycle management across multiple domains. Some works focus on single-domain slice management and control. Chang and Lin~\cite{free5gmano} coordinate OSS/BSS and MANO for NSI-related management in the CN domain. Their Day-0 operations are limited to template-driven preparation of CN slice descriptors, including NSD, VNFD, and NSST design, which are later used to realize the CN part of NSIs during Day-1 operations. Their Day-2 support is partial and mainly targets VNF or VM failure handling. ORANSlice~\cite{oranslice}, in contrast, focuses on the RAN domain by extending OpenAirInterface with 3GPP-compliant RAN slicing, multi-PDU support, and Near-RT RIC-based slice control through an E2SM-CCC service model and a slicing xApp. ORANSlice provides Day-2-like RAN control capabilities, such as runtime PRB-policy adaptation and minimum radio-resource guarantees, but does not provide orchestration-level Day-0 Service Profile modeling, Day-1 end-to-end NSI instantiation, or coordinated RAN--CN QoS enforcement.

Other works target end-to-end NSI lifecycle management by considering both RAN and CN domains. Frameworks such as ONAP~\cite{onap_so_architecture}, CLiSO~\cite{cliso}, and NASP~\cite{nasp} provide broader slice orchestration capabilities than single-domain approaches, but they generally rely on imperative or workflow-driven lifecycle logic and predefined slice templates. As a result, the mapping from application-level requirements to slice-specific configurations often depends on static descriptors or human intervention, rather than being dynamically derived from application-centric Service Profiles. Moreover, coordination between the RAN and CN domains for orchestration and QoS enforcement is not natively part of their lifecycle management for NSIs. In contrast, as shown in Table~\ref{tab:orchestrator-classification}, METIS (a) treats NSIs as first-class logical resources, (b) dynamically derives 3GPP-aligned Slice Profiles rather than relying on static templates, (c) natively coordinates O-RAN and 3GPP slicing across the RAN and CN domains, and (d) supports full Day-0/1/2 lifecycle management with coordinated QoS enforcement. 

%%%% High-level METIS Architecture %%%%%%
\begin{figure}[!t]
    \centering
    \includegraphics[width=\linewidth]{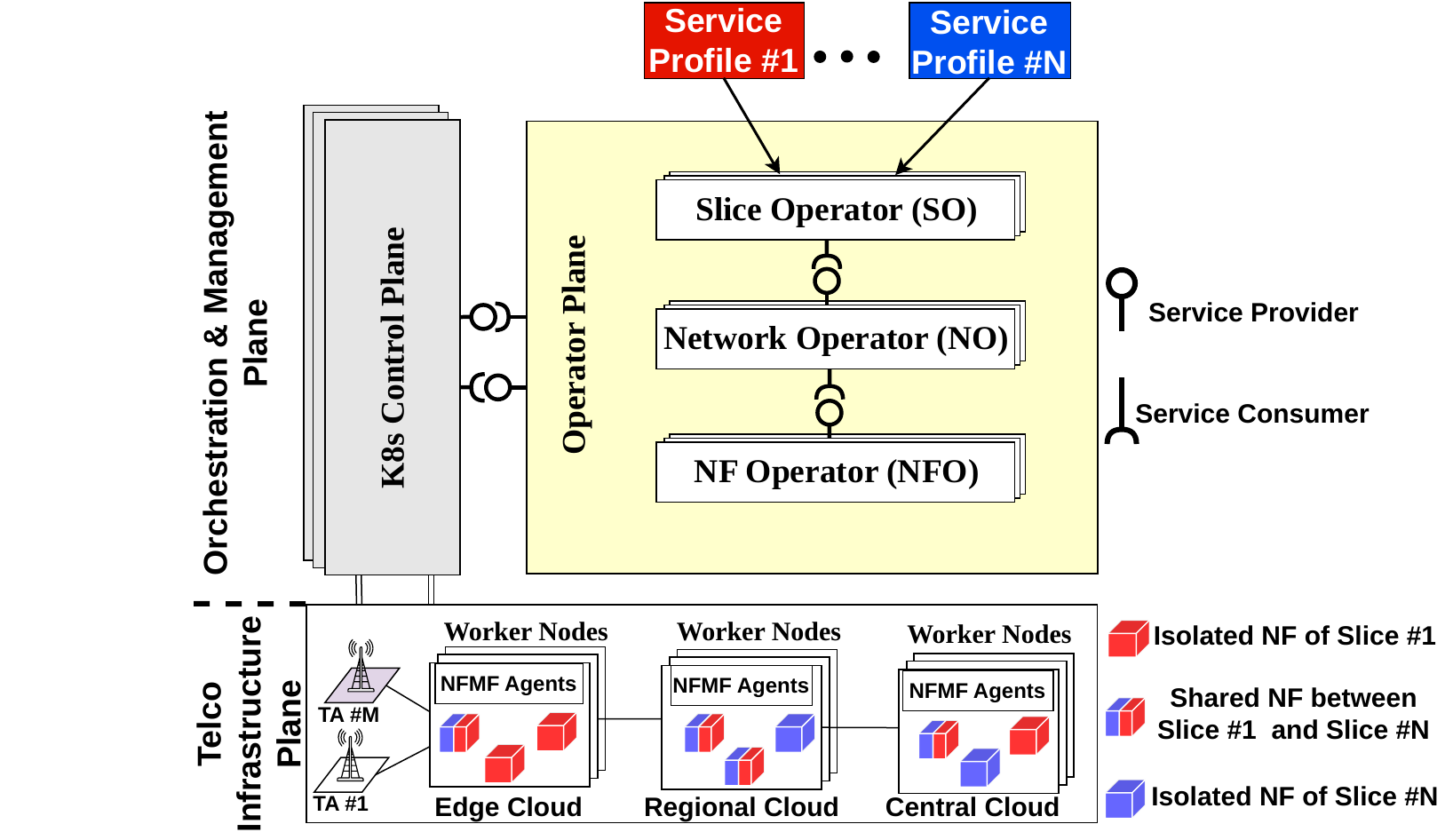}
    \caption{High-level METIS Architecture.}
    \label{fig:higlevel-sliceopeartor-architecture}
\end{figure}
%%%%%%%%%%%%% End %%%%%%%%%%%%%%%%%%%%%

\section{\textsf{Design} Overview} \label{Sec:design_overview}

The principal contribution of this work, \textsc{METIS}, is a declarative and reconciliation-driven network slice orchestrator. It jointly coordinates 3GPP slicing and O-RAN slicing, thereby enabling the dynamic optimization of QoS guarantees across different service categories. Moreover, it addresses an unresolved operational challenge for NSIs in cloud-native environments: the simultaneous lifecycle management of NSIs across both RAN and CN domains while dynamically enforcing NSI-level QoS.
%%%%%

Towards this end, the Orchestration \& Management Plane is distinguished from the Telco Infrastructure Plane (Fig.~\ref{fig:higlevel-sliceopeartor-architecture}). The Telco Infrastructure Plane comprises radio equipment and computational resources (worker nodes across Edge, Regional, and Central Clouds), on which Containerized NFs (CNFs) run. These NFs are configured, started, and terminated by NFMF agents---per-NF agents realizing the Network Function Management Function role defined in 3GPP TS 28.533~\cite{ts28533}---based on general or slice-specific configurations issued by the Orchestration \& Management Plane. In addition, NFMF agents resolve inter-NF dependencies through inter-agent communication; such dependencies arise either from Day-0/1/2 NSI operations or from the functional requirements of the NFs.

The Orchestration \& Management Plane governs all the discussed components in the Telco Infrastructure Plane. It consists of the Kubernetes Control Plane and the Operator Plane. The Operator Plane hosts declarative and reconciliation-driven operators that are designed and developed using the Operator Pattern~\cite{operatorpattern}. The NF Operator, Network Operator, and Slice Operator constitute the Operator Plane, responsible for the lifecycle management of NFs, NSs, and NSIs, respectively. This layering establishes a chain of dependent lifecycles: the lifecycle of NSIs is coupled with that of NSs, which is in turn coupled with that of NFs, which in turn depends on the lifecycle of Pods and worker nodes managed by the Kubernetes Control Plane. Therefore, as shown in Fig.~\ref{fig:higlevel-sliceopeartor-architecture}, the Operator Plane and the Kubernetes Control Plane coordinate by consuming each other's services.

%%%% NSI-NS-NF-DP-Mapping %%%%%%
\begin{figure}[!t]
    \centering
    \includegraphics[width=1\linewidth]{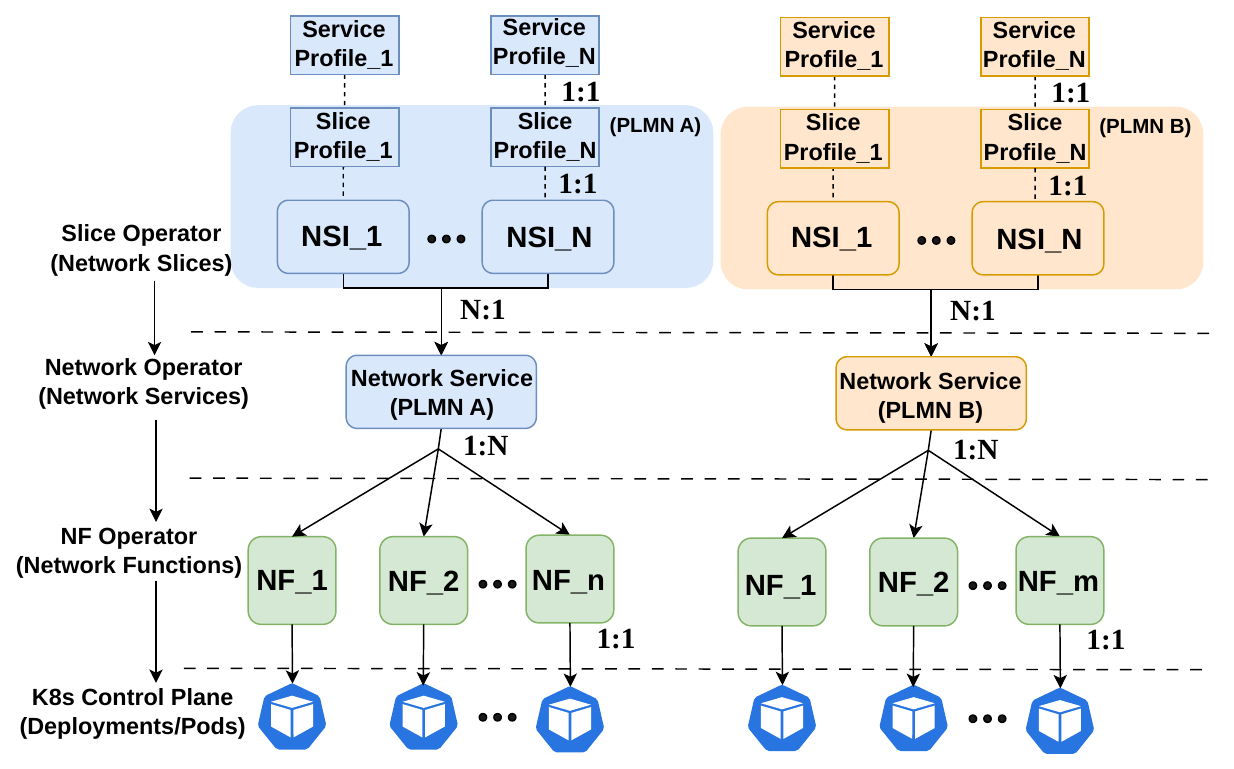}
    \caption{Hierarchical Mapping for NSI Orchestration.}
    \label{fig:hierarchical_mapping_for_nsi_orchestration}
\end{figure}
%%%%%%%%%%%%% End %%%%%%%%%%%%%%%%%%%%%

In METIS, the Slice Operator is the core component that performs the lifecycle management of NSIs across a geographical area by consuming the Network Operator's services. To this end, it creates, updates, and deletes NSs to manage the lifecycle of NSIs. Upon receiving a Service Profile, the Slice Operator derives the corresponding Slice Profile as part of Day-0 operations for NSIs following the 5G QoS model (Subsection~\ref{subsec:day0}), resulting in a one-to-one (1:1) mapping between each Service Profile and each Slice Profile, as shown in Fig.~\ref{fig:hierarchical_mapping_for_nsi_orchestration}. In METIS, each Slice Profile describes the specification of one NSI that should be established by mapping its components to the appropriate NFs within the NS. The Slice Operator establishes this mapping following the slicing and scaling principle: it slices existing NFs within the NS across NSIs when sharing is permitted, and otherwise scales out new NFs subject to admission control (e.g., available resources). To elaborate, if the NS already contains a suitable NF that can host the relevant component of the NSI, that NF is sliced by assigning the NSI's slice identifier and resources to it. Otherwise, the Slice Operator scales out one or more new NFs into the NS---e.g., when existing NFs are dedicated to other NSIs, no NF is deployed in the requested geographical area, or the deployed NFs serve a different slice/service type (see Subsection~\ref{subsubsection:nsi_instantiation}). Moreover, the Slice Operator determines which NFs in the NS should be connected to one another by specifying the NSI serving scope for each NF, which is later used by the NFMF agents to configure each NF with its serving NSIs. Since the NSIs of other Slice Profiles are established in the same NS, this results in a many-to-one (N:1) mapping from NSIs to NS. Furthermore, as can be seen in Fig.~\ref{fig:hierarchical_mapping_for_nsi_orchestration}, the Slice Operator maintains a separate NS for each Mobile Network Operator (MNO), identified by its Public Land Mobile Network (PLMN), and the NSIs of an MNO are mapped to its associated NS, which yields a one-to-one (1:1) mapping between each PLMN and its NS\footnote{For simplicity, we assume that each MNO operates on dedicated infrastructure, i.e., no infrastructure is shared among MNOs.}.

%%%% Slice Operator Architecture %%%%%%
\begin{figure}[!t]
    \centering
    \includegraphics[width=1\linewidth]{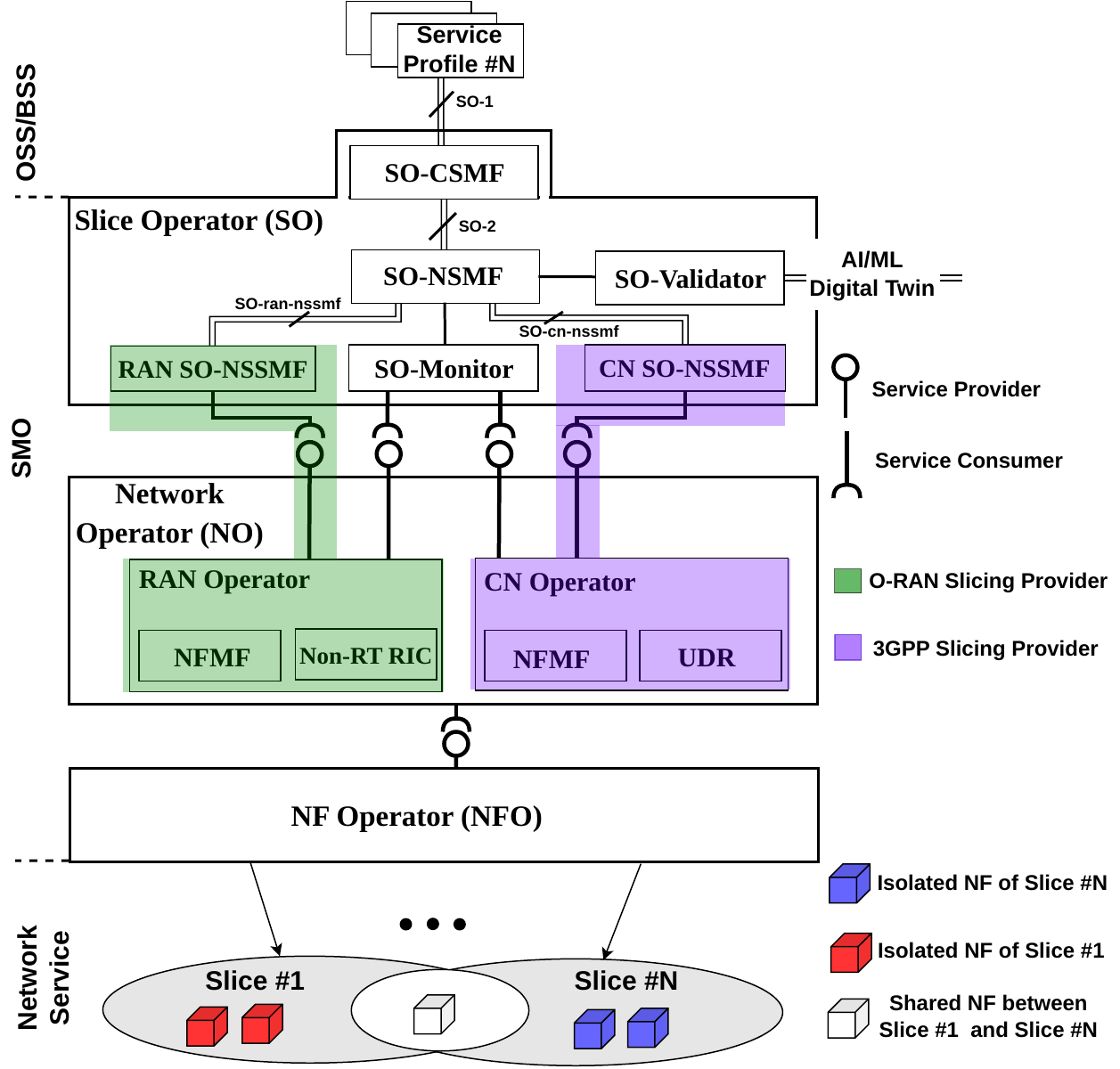}
    \caption{METIS Detailed Architecture.}
    \vspace{-1.0em}
    \label{fig:metis-detailed-architecture}
\end{figure}
%%%%%%%%%%% End %%%%%%%%%%%%%%%%%%%%%%%

Upon receiving the specification of an NS determined by the Slice Operator, the Network Operator creates, updates, or deletes the specifications of the NFs as input to the NF Operator, which results in a one-to-many (1:N) mapping between each NS and its constituent NFs. Based on the action defined for an NF---creation, update, or deletion---the NF Operator performs the lifecycle management of the corresponding Deployment (a native Kubernetes resource), maintaining a one-to-one (1:1) mapping between each NF and its associated Deployment.

All operations involved in this hierarchical mapping---whether for NSI establishment or for subsequent updates, upgrades, and deletions---are designed to be idempotent, ensuring consistent state convergence under repeated executions (see Subsection~\ref{subsubsection:day-2}). This property is particularly critical in scenarios where NSIs share underlying NFs, as lifecycle actions on one NSI must not adversely affect others. To guarantee this, the reconciliation loops of the Slice Operator and Network Operator enforce idempotent control actions through an observe–compare–act paradigm, ensuring that only the necessary changes are applied by observing the actual state of each NSI, comparing it with its desired state, and acting to make them converge.

\section{SYSTEM DESIGN \& IMPLEMENTATION} \label{Sec:systemdesign_and_implementatoin}

This section presents the system design and implementation of METIS, detailing the key components of the Slice Operator, Network Operator, and NF Operator, as well as their interactions for full lifecycle management of NSIs. It further describes how Day-0 (Service Modeling and NSI Design), Day-1 (NSI Instantiation and Resource Orchestration), and Day-2 (NSI Runtime Control and Lifecycle Operations) operations are realized through coordinated management, control, and monitoring mechanisms.

\subsection{Service Modeling \& NSI Design (Day-0)} \label{subsec:day0} %% done

The highest level of abstraction in METIS, represented by the first Day-0 operation in the Slice Operator, is realized by transforming application-centric Service Profiles into 3GPP-aligned Slice Profiles. The data model for application-centric Service Profiles introduced by METIS comprises semantic information and QoE requirements. The semantic information defines the application-level meaning of the service traffic, including the use-case type, target data networks, DNS endpoints, traffic type and pattern, application and transport protocols, port information, and payload/content types. The QoE requirements specify the expected performance and deployment characteristics, including content rates, traffic class, isolation requirements, latency class, reporting period, coverage region and zone, user density, and user-equipment type \footnote{An example of an
Application-centric Service Profile is publicly available in the
\url{https://bubbleran.com/docs/v3.0.0/user-guide/slice-training/lab02}
(accessed: Jul.\ 6, 2026).} . CSCs express the network requirements of their applications in these application-centric Service Profiles, and using the SO-1 interface (Fig.~\ref{fig:metis-detailed-architecture}), they submit them to the SO-CSMF.

%%%% Slice PDUSession Flow Mapping %%%%%%
\begin{figure}[!t]
    \centering
    \begin{minipage}[t]{0.48\linewidth}
        \centering
        \includegraphics[width=\linewidth]{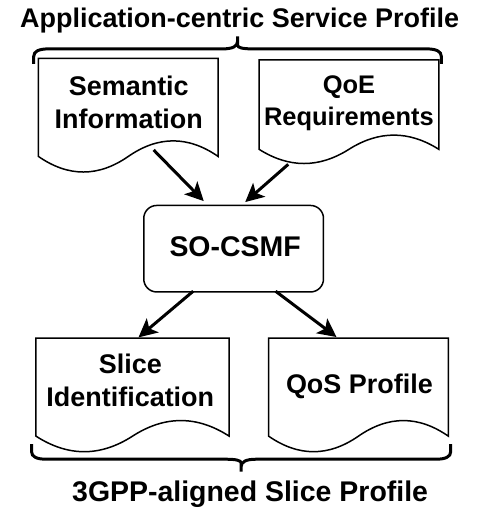}
        \caption{Service Profile to Slice Profile derivation by the SO-CSMF.}
        \label{fig:serviceprofile-to-sliceprofile}
    \end{minipage}\hfill
    \begin{minipage}[t]{0.48\linewidth}
        \centering
        \includegraphics[width=\linewidth]{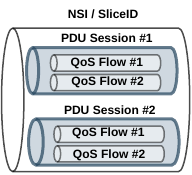}
        \caption{Hierarchical relationship among NSI, PDU Session, and Flow.}
        \label{fig:nsi-pdusession-flow-mapping}
    \end{minipage}
\end{figure}

% \begin{figure}[!t]
%     \centering
%     \includegraphics[width=0.4\linewidth]{Figs/Hierarchical_Qos_Model .pdf}
%     \caption{Hierarchical relationship among NSI, PDU Session, and Flow.}
%     \label{fig:nsi-pdusession-flow-mapping}
% \end{figure}
%%%%%%%%%%%%% End %%%%%%%%%%%%%%%%%%%%%

To design an NSI, the SO-CSMF derives its 3GPP-aligned Slice Profile from the corresponding application-centric Service Profile as shown in Fig.~\ref{fig:serviceprofile-to-sliceprofile}. The data model of a Slice Profile consists of slice identification attributes and QoS profile based on 5G QoS model in TS 23.501~\cite{ts23501}. The slice identification attributes, composed of PLMN, S-NSSAI, and SliceID, are determined based on the semantic information of Service Profiles. For instance, based on the defined use-case category (e.g., online gaming, data transfer), the SST (Slice/Service Type) value of an S-NSSAI is specified. Then, to ensure uniqueness, the SD (Slice Differentiator) value is calculated based on the number of existing Slice Profiles with the same SST value. Finally, the SO-CSMF calculates the numeric value of the corresponding SliceID as $\text{SliceID} = (\text{SST} \ll 24) \mathbin{|} \text{SD}$, in which the SST occupies the 8 most significant bits and the SD occupies the 24 least significant bits.

To derive the QoS profile of a Slice Profile, the SO-CSMF employs a bottom-up hierarchical aggregation approach that is aligned with the granularity of the 5G QoS model. As illustrated in Fig.~\ref{fig:nsi-pdusession-flow-mapping}, an NSI can support multiple PDU sessions, each associated with a DNN, and each PDU session may comprise one or more QoS Flows, each identified by a QFI. Accordingly, the SO-CSMF starts from the QoS Flow level, the finest granularity, and progressively aggregates QoS requirements toward the PDU session level and, finally, the NSI level, the coarsest granularity. This process uses both the semantic information and the QoE requirements contained in the corresponding Service Profiles. To this end, the SO-CSMF uses the semantic information, particularly the traffic type of each data flow within the data networks, to determine the QFI and the corresponding QoS Flow parameters (e.g., ARP (Allocation and Retention Priority), PLR (Packet Loss Rate), GBR (Guaranteed Bit Rate), and MBR (Maximum Bit Rate)) through a traffic-type-to-QoS model inspired by Table 5.7.4-1 of 3GPP TS 23.501~\cite{ts23501}. In parallel, the SO-CSMF uses the QoE requirements, especially the content rate, to adjust the GBR and MBR parameters of the corresponding QoS Flows. It then determines the AMBR (Aggregate Maximum Bit Rate) of each PDU session by aggregating the requirements of one or more QoS Flows associated with that PDU session. Ultimately, by aggregating the AMBR values across PDU sessions, the SO-CSMF determines the NSI AMBR, which represents the NSI-level QoS \footnote{An example of a 3GPP-defined Slice Profile is publicly available in the
\url{https://bubbleran.com/docs/v3.0.0/user-guide/slice-training/lab01}
(accessed: Jul.\ 6, 2026).}.

\subsection{NSI Instantiation \& Resource Orchestration (Day-1)} \label{subsubsection:nsi_instantiation} %%% done

To jointly realize O-RAN and 3GPP slicing across the RAN and CN domains, METIS introduces an O-RAN slicing provider (comprising the RAN SO-NSSMF and the RAN Operator) and a 3GPP slicing provider (comprising the CN SO-NSSMF and the CN Operator), as shown in Fig.~\ref{fig:metis-detailed-architecture}. Acting as the coordinator, the SO-NSMF declaratively drives the RAN SO-NSSMF and the CN SO-NSSMF to instantiate NSIs and orchestrate resources across both domains according to the Slice Profiles issued in Day-0 by the SO-CSMF. To trigger the Day-1 operations, the SO-NSMF derives domain-specific Slice Profiles, namely RAN Slice Profiles and CN Slice Profiles, from these Slice Profiles and delegates them to the corresponding SO-NSSMFs. The RAN Slice Profile contains low-level parameters required to create the NSI in the RAN domain, such as PLMN, S-NSSAI, coverage area, user density, isolation requirements, UE type, and Slice AMBR. Similarly, the CN Slice Profile contains low-level parameters required to create the NSI in the CN domain, such as PLMN, S-NSSAI, DNN, PDU session type (IPv4 or IPv6), Slice AMBR, session AMBR, isolation requirements, and QoS flow parameters.

In the RAN domain, the RAN SO-NSSMF instantiates an NSI by determining whether the O-RAN NFs within the requested geographical coverage area should be scaled out or sliced. To this end, it first checks whether suitable O-RAN NFs (i.e., those already deployed with sufficient PRBs) are available to host the requested NSI in the target area, which is mapped to a Tracking Area (TA). If so, it instructs the RAN Operator to slice the selected O-RAN NFs by assigning the S-NSSAI of the NSI to them. Otherwise, if sufficient radio and computational resources exist (as declared in the Resource Inventory Profiles) to deploy or expand the required O-RAN NFs, it instructs the RAN Operator to scale them out and assign the NSI to them.

Similarly, in the CN domain, the CN SO-NSSMF instantiates an NSI by determining whether the 5G CN NFs associated with the data plane or control plane should be scaled out or sliced. The CN SO-NSSMF takes this decision based on the SST, isolation requirements, latency class, already-deployed 5G CN NFs, 5G CN characteristics, and the computational resources available in the target area. Specifically, if an NSI does not require an isolated 5G CN control plane, the CN SO-NSSMF instructs the CN Operator to slice the currently deployed 5G CN control-plane NFs by assigning the NSI to them. As another case, if the latency class of the NSI is MEC (Mobile Edge Computing) and no 5G CN data-plane NFs with the same SST have been deployed in the target area, the CN SO-NSSMF instructs the CN Operator to scale out 5G CN data-plane NFs, provided that sufficient resources are available.

Instantiating NSIs by adding or updating O-RAN/5G NFs in the associated NS does not by itself define the NF interconnection topology. To this end, the RAN SO-NSSMF and CN SO-NSSMF declaratively specify the NSI serving scope of NFs. Accordingly, two or more NFs are interconnected when: (1) their NSI serving scopes overlap, and (2) a corresponding O-RAN interface or 3GPP reference point exists between them. This enables both SO-NSSMFs to abstractly specify the inter-NF topology without directly managing low-level configurations, which are delegated to the NFMF within the RAN and CN Operators.

Regarding admission control and resource allocation for NSIs, METIS does not prescribe a specific optimization algorithm; rather, it defines the SO-Validator as a declarative extension point attached to the SO-NSMF, through which external AI/ML- or Digital-Twin-based mechanisms can plug in their optimization decisions. While remaining agnostic to the optimization mechanisms, the SO-NSMF exposes necessary information such as Slice Profiles, the current NS, and monitoring information through this interface. These decisions indicate whether a requested NSI can be admitted while satisfying the slicing or scaling rules for O-RAN/5G NFs described above, and would be consumed by the SO-NSMF, RAN SO-NSSMF, and CN SO-NSSMF during the Day-1 operations.

\subsection{NSI Runtime Control \& Lifecycle Operations (Day-2)} \label{subsubsection:day-2} %%% done

Day-2 operations are where the effectiveness of declarative and reconciliation-driven logic for NSI orchestration becomes evident. Once instantiated, an NSI must remain aligned with the evolving requirements of the applications or services it serves throughout its operational lifetime. METIS achieves this by propagating the NSI orchestration logic across the cascaded reconciliation loops in a multi-level manner, as shown in Fig.~\ref{fig:metis-detailed-architecture}: the outputs of upper-level components (target states) are cascaded as desired states for lower-level reconciliation loops, thereby enabling idempotent lifecycle management for NSIs. The following sub-subsections describe how METIS realizes Day-2 monitoring, QoS enforcement, NSI modification, and NSI termination.

\subsubsection{NSI Monitoring}

The SO-Monitor, within the Slice Operator (Fig.~\ref{fig:metis-detailed-architecture}), collects NSI-level metrics and derives the actual state of each NSI with respect to lifecycle progress and performance. In particular, it determines whether an NSI has been created, modified, or deleted by computing the lifecycle management metrics, as defined in Table~\ref{tab:cloudnative_lcm_metrics}. In addition, it evaluates the performance of each NSI by assessing whether the QoE requirements requested by CSCs are satisfied, according to the SLA ratio defined in Eq.~\ref{eq:sla_formula}. The SO-Monitor then updates the status of the NSIs, enabling the SO-NSMF reconciliation loop to detect deviations between the desired and actual states and trigger the required corrective actions.

\subsubsection{NSI-level QoS Enforcement} \label{subsubsection:qos_enforcement}

In METIS, QoS enforcement is treated as a built-in part of the NSI lifecycle and is performed once an NSI has been instantiated across the RAN and CN domains. Unlike typical O-RAN control operations, which optimize radio resources independently of CN QoS policies, METIS harmonizes O-RAN and 3GPP control operations to enforce NSI-level QoS across both domains. To this end, the SO-NSMF coordinates enforcement by instructing the RAN SO-NSSMF and the CN SO-NSSMF to enforce NSI-level QoS in their respective domains. The following paragraphs detail this enforcement.

The RAN SO-NSSMF derives the RAN SLA assurance requirements---including slice identification, the target Near-RT RIC, the scope of the monolithic or disaggregated O-gNB, and the NSI-level QoS---from the RAN Slice Profile and the list of O-RAN NFs serving the NSI. It then uses the declarative R1 interface to convey the RAN SLA assurance requirements to the Non-RT RIC within the RAN Operator (Fig.~\ref{fig:metis-detailed-architecture}). Upon receiving the SLA assurance requirements, the Non-RT RIC forwards them to the target Near-RT RIC over the A1 interface, where the SLA xApp is deployed to process and enforce them. To this end, the xApp subscribes to E2SM-KPM reports for monitoring and invokes E2SM-CCC control actions over the E2 interface between the Near-RT RIC and the target O-gNB. Once the required E2 procedures are established, the SLA xApp operates in a closed observe-compare-act loop to dynamically adjust PRB allocation for the NSI: if the allocated PRBs are insufficient to satisfy the NSI-level QoS, it increases the allocation; otherwise, it decreases the allocation to constrain NSI traffic to the requested QoS.

The CN SO-NSSMF derives the slice subscription information, including slice identification and QoS rules, from the corresponding CN Slice Profile and stores it in the UDR serving that NSI for use by the PCF. To enforce NSI-level QoS, each CN data-plane NF is paired with a Traffic-Shaper (realized as a sidecar workload; see Section~\ref{sec:implementation}), to which the CN SO-NSSMF provides the NSI-level QoS configuration. The Traffic-Shaper then constrains the traffic of an NSI in the CN domain according to its QoS requirements by placing the downlink and uplink traffic into two separate queues and regulating their rates at the CN data plane based on the NSI-level QoS (Slice AMBR).

\subsubsection{NSI Modification (Update and Upgrade)} \label{subsubsection:nsi_modification}
CSCs may update their Service Profiles during Day-2 operations to reflect the evolving requirements of their customers, applications, or services. The SO-CSMF observes these changes in Service Profiles and modifies the corresponding Slice Profiles accordingly. The modified Slice Profiles represent a new desired state for the associated NSIs.

However, not every Slice Profile modification requires the same type of operation on the corresponding NSI. If the modified parameters only affect the runtime QoS behavior of the NSI, the SO-NSMF treats the modification as an NSI update. In this case, only control operations are required to re-enforce the requested QoS in the RAN and CN domains, which is similar to NSI-level QoS enforcement.

In contrast, if the modified Slice Profile changes structural properties of an NSI, the SO-NSMF identifies the modification as an NSI upgrade. Such structural properties include the set of NFs in the NS, their placement, their sharing model, or their interconnection topology. In this case, the currently deployed NSI can no longer satisfy the new desired state through re-enforcing NSI-level QoS alone. Specifically, if an NSI's coverage area changes entirely, the existing NSI is no longer valid for the requested geographical scope and must be terminated, re-onboarded, and re-instantiated to realize the upgrade.

For the above reasons, an NSI upgrade requires multiple reconciliation steps. First, during an upgrade, the SO-NSMF reconciles the system toward a state in which the old NSI is no longer associated with the modified Slice Profile by initiating the termination of the old NSI (see Sub-subsection~\ref{subsubsection:nsi_termination}). Second, it reconciles the system toward the onboarding and instantiation of the new NSI based on the modified RAN and CN Slice Profiles. Finally, it reconciles the system toward enforcing QoS for the new NSI.

\subsubsection{NSI Termination and Resource Release} \label{subsubsection:nsi_termination}

When a CSC no longer requires an NSI, it can request termination by deleting the corresponding Service Profile. The SO-CSMF detects this deletion and removes the corresponding Slice Profile, which represents the termination of the associated NSI as the new desired state. The SO-NSMF identifies this and marks the NSI as dangling (i.e., its Slice Profile no longer exists). To propagate this new desired state, the SO-NSMF deletes the RAN and CN Slice Profiles derived from the Slice Profile, which triggers the RAN SO-NSSMF and CN SO-NSSMF to terminate the NSI within their respective domains.

Terminating a dangling NSI may require updating or deleting the O-RAN/5G NFs and their interconnections in the associated NS. If one or more NFs exclusively serve the dangling NSI, the corresponding SO-NSSMF deletes those NFs. Otherwise, for shared NFs, the SO-NSSMF updates their annotations and NSI serving scopes by removing the identifiers of the dangling NSI. This prevents the termination of one NSI from disrupting other NSIs that share the same underlying NFs. However, the mentioned operations alone are not sufficient, because any RAN and CN resources or configurations allocated during QoS enforcement must also be released. For instance, the CN SO-NSSMF removes the slice subscription information of the dangling NSI from the relevant UDR or database. Similarly, the RAN SO-NSSMF releases the PRBs allocated to the dangling NSI by deleting its SLA assurance requirements over the R1 interface of the Non-RT RIC, which implicitly triggers the SLA xApp to release them. After these management and control operations are completed, the dangling NSI is safely removed without affecting other active NSIs.

\subsection{Implementation Details} \label{sec:implementation}

We implemented a prototype of METIS that follows the system design described above. All Slice Operator components were developed from scratch in Go (v1.22) using the Operator SDK~\cite{operatorsdk}. For the Network Operator and NF Operator, we reused two components from Athena~\cite{athena}, namely the Base Operator and the Manager, to orchestrate and configure O-RAN/5G NFs. In addition, we implemented the Non-RT RIC from scratch in Go using the Operator SDK to support the QoS enforcement in the RAN domain as part of lifecycle management through control operations.

For the O-RAN/5G NFs, we built container images based on OpenAirInterface (OAI)~\cite{oai} and Open5GS~\cite{open5gs}, respectively, and stored them in a private container registry for reuse throughout the lifecycle of NSIs. The Traffic-Shaper and SLA xApp, which are used for QoS enforcement, were also developed from scratch in Go and C, respectively, and containerized to automate their deployment. Overall, the METIS prototype includes approximately 20K lines of code for the components developed from scratch. The Traffic-Shaper runs as a sidecar workload alongside the UPF workload, whereas the SLA xApp runs as a separate workload in its own Pod and connects to FlexRIC~\cite{flexric}, a Near-RT RIC implementation, through the E42 interface. In the following, we describe the implementation details of NSI lifecycle management.

\textbf{Day-0 Operations.} In our implementation, the logic of the SO-CSMF is realized by a Service Profile Controller, a custom Kubernetes controller. Through the Kubernetes API Server, the Service Profile Controller watches \texttt{ServiceProfile} resources defined by CSCs and transforms each into a \texttt{SliceEntity}---the implementation-level representation of a Slice Profile---inside a \texttt{Slice} object that serves as the internal desired-state model consumed by the Slice Controller for Day-1 and Day-2 operations. During reconciliation, the Service Profile Controller creates the \texttt{Slice} object if it does not already exist, then computes the corresponding \texttt{SliceEntity} and adds it to the \texttt{Slice} object. To support idempotent reconciliation and later Day-2 modifications, the Service Profile Controller stores the last-applied Service Profile specification as a JSON annotation and updates the \texttt{SliceEntity} only when the desired state changes. Finally, it attaches a finalizer to each Service Profile, ensuring that deletion first removes the associated \texttt{SliceEntity} from the \texttt{Slice} object before the resource is garbage-collected.

\textbf{Day-1 Operations.} In our implementation, the SO-NSMF, RAN SO-NSSMF, CN SO-NSSMF, and SO-Monitor are jointly realized by the Slice Controller, another custom Kubernetes controller. It observes the \texttt{Slice} object created during Day-0 operations and maps its desired state into a \texttt{Network} object, the implementation-level representation of an NS. The Slice Controller maps each \texttt{SliceEntity} to the O-RAN/5G NFs required to instantiate the corresponding NSI. Slicing and scaling decisions are reflected either by assigning an NSI to suitable shared NFs or by adding new NF instances when isolation or capacity requirements demand it. The resulting \texttt{Network} object is then handled by the reused Base Operator, which decomposes the network-level desired state into NF-level \texttt{Element} objects, one per NF, and deploys the corresponding containerized workloads. Slice-specific configurations are issued through the reused Manager, enabling the configuration of inter-NF connectivity and assigning each NSI to the appropriate O-RAN/5G NFs. Finally, a slice-level finalizer ensures that changes in slice membership are propagated to the \texttt{Network} object before the \texttt{Slice} resource is reclaimed.

\textbf{Day-2 Operations.} In our implementation, Day-2 operations are mainly handled by the Service Profile Controller and the Slice Controller. The Service Profile Controller detects changes in \texttt{ServiceProfile} objects by comparing the current specification with the last-applied specification stored as an annotation. If the Service Profile Controller detects an NSI update, it updates the QoS parameters of the corresponding \texttt{SliceEntity} in the \texttt{Slice} object. Otherwise, if it detects an NSI upgrade, it deletes the old \texttt{SliceEntity} from the \texttt{Slice} object and adds the new one so that the Slice Controller deletes the old NSI and recreates a new one. Regarding NSI monitoring, the Slice Controller reconciles the \texttt{Slice} object every minute so that \texttt{SliceMonitor}, the implementation-level representation of the SO-Monitor, determines the status of each \texttt{SliceEntity} by retrieving the status of the Pods, Deployments, and \texttt{Elements} from the Kubernetes control plane. The SliceID is used as an identifier to filter and classify the information of these components for each \texttt{SliceEntity}. With the SliceID, \texttt{SliceMonitor} also retrieves the performance metrics of \texttt{SliceEntities} from the time-series database stored by the SLA xApp and the Traffic-Shaper. For the QoS enforcement, the Traffic-Shaper generates the Linux Traffic Control rules~\cite{tc} based on the NSI-level QoS (Slice AMBR) and binds them to the tunnel network interface (TUN) in UPF that serves an NSI. Additionally, it captures the performance metrics to be collected by Prometheus. Finally, NSI termination is implemented through finalizers across the \texttt{ServiceProfile}, \texttt{Slice}, \texttt{Network}, and \texttt{Element} resources, ensuring that slice membership, NF configuration, and associated workloads are released in the correct hierarchical order.

\begin{table}[t]
\centering
\footnotesize
\setlength{\tabcolsep}{3pt}
\caption{Specifications of Traffic Types in the Campus Event.}
\label{tab:campus-event-use-case}
\begin{tabular}{@{}lll@{}}
\toprule
\shortstack[l]{\textbf{Service}\\\textbf{Profile}} & \shortstack[l]{\textbf{Use Case}\\\textbf{Type}} & \textbf{QoE Requirement} \\
\midrule
\multirow{4}{*}{\shortstack[l]{Security\\Cameras}}
& \multirow{4}{*}{\shortstack[l]{Video\\Streaming}}
&   MAX (DL: 12\,Mbps, UL: 12\,Mbps) \\
& & AVG (DL: 10\,Mbps, UL: 10\,Mbps) \\
& & GTD (DL: 8\,Mbps, UL: 8\,Mbps) \\
& & TCT: 80\% \\
\addlinespace
\multirow{4}{*}{Organizers}
& \multirow{4}{*}{Video Chat}
&   MAX (DL: 60\,Mbps, UL: 9.6\,Mbps) \\
& & AVG (DL: 50\,Mbps, UL: 8\,Mbps) \\
& & GTD (DL: 40\,Mbps, UL: 6.4\,Mbps) \\
& & TCT: 80\% \\
\addlinespace
\multirow{4}{*}{Participants}
& \multirow{4}{*}{\shortstack[l]{Data\\Transfer}}
&   MAX (DL: 48\,Mbps, UL: 3.6\,Mbps) \\
& & AVG (DL: 40\,Mbps, UL: 3\,Mbps) \\
& & GTD (DL: 32\,Mbps, UL: 2.4\,Mbps) \\
& & TCT: 80\% \\
\bottomrule
\end{tabular}
\vspace{2pt}

\parbox{\columnwidth}{\raggedright\footnotesize
MAX: Maximum; AVG: Average; GTD: Guaranteed; TCT: Time Compliance Threshold.}
\end{table}

% \begin{table}[t]
% \centering
% \caption{Specifications of Traffic Types in the Campus Event}
% \label{tab:campus-event-use-case}
% \begin{tabular}{p{1.8cm} p{1.8cm} p{4cm}}
% \toprule
% \textbf{Service Profile} & \textbf{Use Case Type} & \textbf{QoE Requirement} \\
% \midrule
% \multirow{4}{*}{{Security Cameras}} 
% & \multirow{4}{*}{Video Streaming} 
% &   MAX (DL: 12Mbps, UL: 12Mbps) \\
% & & AVG (DL: 10Mbps, UL: 10Mbps) \\
% & & GTD (DL: 8Mbps, UL: 8Mbps) \\
% & & TCT: 80\% \\
% \addlinespace
% \multirow{4}{*}{{Organizers}} 
% & \multirow{4}{*}{Video Chat} 
% &   MAX (DL: 60Mbps, UL: 9.6Mbps) \\
% & & AVG (DL: 50Mbps, UL: 8Mbps) \\
% & & GTD (DL: 40Mbps, UL: 6.4Mbps) \\
% & & TCT: 80\% \\
% \addlinespace
% \multirow{4}{*}{{Participants}} 
% & \multirow{4}{*}{Data Transfer} 
% &   MAX (DL: 48Mbps, UL: 3.6Mbps) \\
% & & AVG (DL: 40Mbps, UL: 3Mbps) \\
% & & GTD (DL: 32Mbps, UL: 2.4Mbps) \\
% & & TCT: 80\% \\
% \bottomrule
% \end{tabular}
% \vspace{0.5ex}
% \footnotesize{ 
% MAX: Maximum; 
% AVG: Average; 
% GTD: Guaranteed; 
% TCT: Time Compliance Threshold.
% }
% \end{table}
%%%%%%%%%%%%%%%%%%%%%%%%%%%%%%%%%
%%%%%%%%%%%%%%%%%%%%%%%%%%%%%%%%

%%%%%%%%%%%%%%%%%%%%%%%%%%%%%%%%%%%%%%%%%%%%%
%%%%%%% Hardware Specification of the Testbed
%%%%%%%%%%%%%%%%%%%%%%%%%%%%%%%%%%%%%%%%%%%%%
\begin{table}[!t]
\centering
\caption{Hardware Specification of the Testbed}
\label{tab:k8s_cluster}
\footnotesize
\setlength{\tabcolsep}{2.5pt}
\renewcommand{\arraystretch}{1.05}
\begin{tabular}{@{}ccccc@{}}
\toprule
\textbf{Node} & \textbf{OS} & \textbf{CPU} & \textbf{RAM} & \textbf{Equipment}\\
\midrule
CP & U22.04 & i9-10920X & 64\,GB & --\\
Worker~1 & U22.04 & i9-10920X & 64\,GB & QTL~RM520NGL (UE)\\
Worker~2 & U22.04 & Ryzen~9950X & 64\,GB & USRP~B210 (RU)\\
Worker~3 & U22.04 & i9-10980XE & 32\,GB & QTL~RM520NGL (UE)\\
Worker~4 & U22.04 & i9-10980XE & 32\,GB & QTL~RM520NGL (UE)\\
\bottomrule
\end{tabular}
\vspace{0.5ex}

\footnotesize{
CP: Control Plane;
UE: User Equipment;
RU: Radio Unit;
QTL: Quectel.
}
\end{table}

%%%%%%%%%%%%%%%%%%%%
%%%%%%%%%%%%%%%%%%%%

\section{Evaluation \& Use Case} \label{Sec:evaluation}

This section evaluates METIS in terms of (1) network slice lifecycle management, (2) joint RAN and CN slicing with QoS enforcement, (3) scalability, and (4) failure recovery. Results for the first two are obtained from a realistic use case, \textbf{Campus Event}, deployed on a 5G cloud-native testbed with an over-the-air interface. For scalability, RAN instances instead run in RF-simulated mode on the same testbed, spanning multiple regions and zones. For failure recovery, faults are injected at four levels---O-RAN NF, CN NF, NS, and NSI---demonstrating how the cascaded reconciliation loops of METIS recover these resources through the observe--compare--act paradigm.

\begin{figure}[!t]
  \centering
  \includegraphics[width=0.5\textwidth]{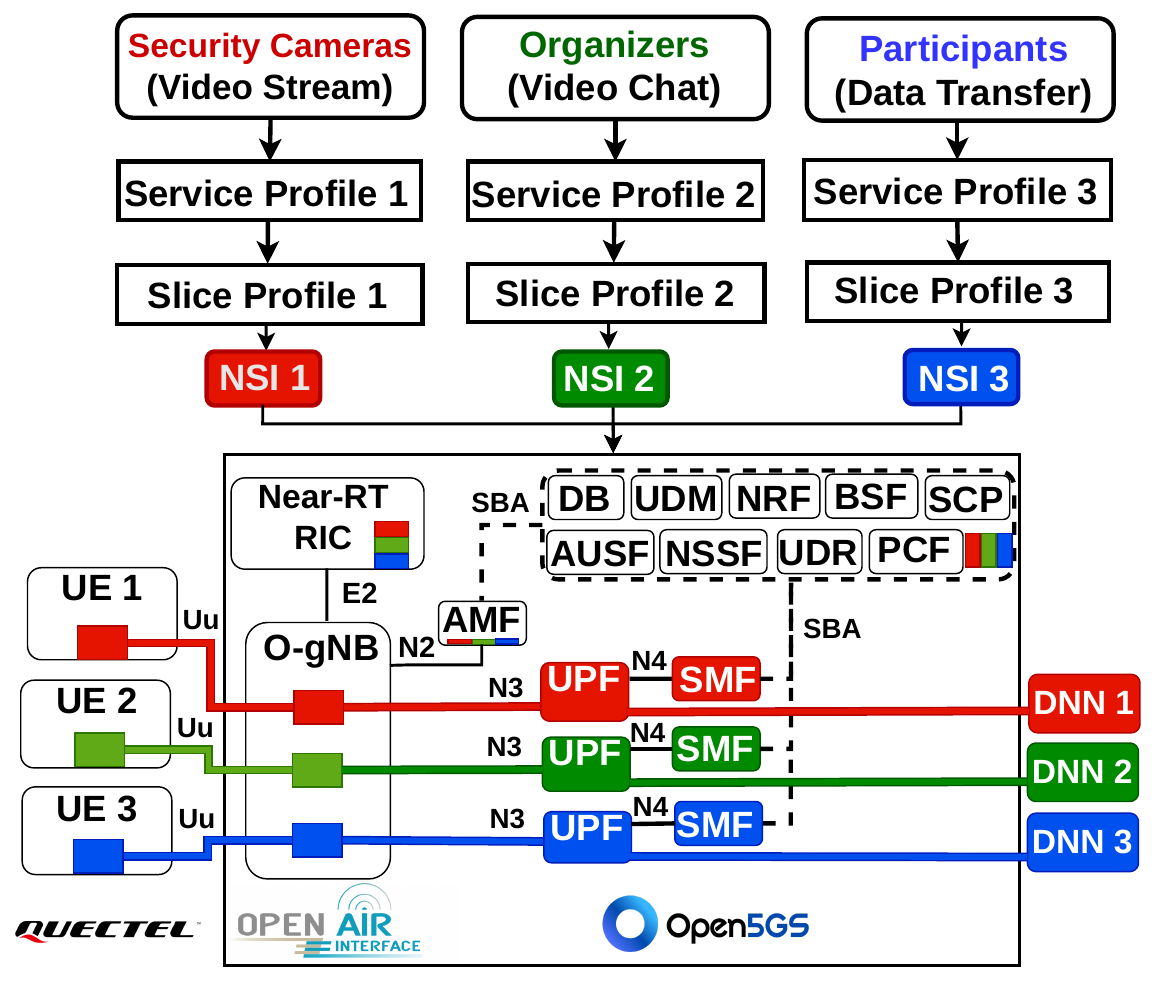}
  \caption{Deployed NSIs for the Campus Event by METIS.}
  \label{fig:campus_event_deploy}
  \vspace{-1em}
\end{figure}

As shown in Table~\ref{tab:campus-event-use-case}, the campus event comprises three traffic categories with distinct QoE requirements and application use-case types: (1) security cameras, involving video streaming from cameras to the event’s security room; (2) organizers, involving peer-to-peer video chat among the event’s organizers; and (3) participants, involving data transfer traffic for the event’s participants. Accordingly, three distinct Service Profiles are defined to capture the network requirements of each traffic type. Fig.~\ref{fig:campus_event_deploy} shows the realization of NSIs when METIS receives three Service Profiles, automatically derives the corresponding Slice Profiles, and instantiates the NSIs by slicing and scaling O-RAN/5G NFs, exemplifying the hierarchical mapping concept of Fig.~\ref{fig:hierarchical_mapping_for_nsi_orchestration} in Section~\ref{Sec:design_overview}. In this deployment, the O-RAN NFs (O-gNB and Near-RT RIC) and all 5G CN NFs except the SMF and UPF are shared across the three NSIs, whereas a dedicated SMF and UPF are instantiated per NSI as isolated NFs. To validate the created NSIs, three Quectel modules acting as real UEs connect to their respective NSIs and exchange uplink and downlink traffic.

For the testbed, a 5G cloud-native cluster is built using Kubernetes (v1.31.3), and the hardware specifications are summarized in Table~\ref{tab:k8s_cluster}. Additionally, Cilium (v1.18.2) is employed as the Container Network Interface (CNI) to enable network connectivity among CNFs, while Containerd (v1.7.27) is installed on each node to support CNF deployment. To collect time-series metrics, Prometheus (v2.38.0) is used. OpenAirInterface5G (v2.3.0) and Open5GS (v2.7.5) are used to deploy the RAN and CN, respectively.
%%%%%%%%%%%%%%%%%%%%%%%%
%%%%%%%%%%%%%%%%%%%%%%%%
%%%%%%%%%%%%%%%%%%%%%%%%
%%%%%%%%%%%%%%%%%%%%%%%%%

\subsection{Network Slice Lifecycle Management} \label{subsection:nsi_lifecycle_result}

\begin{figure*}[!t]
  \centering
  \includegraphics[width=0.8\textwidth]{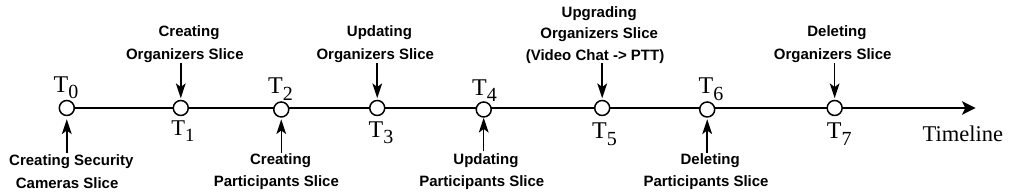}
  \caption{Lifecycle Management Storyline for the Campus Event.}
  \label{fig:networkslice_lcm_timeline}
\end{figure*}

To evaluate the full lifecycle management of NSIs, a representative and realistic scenario based on the campus event is considered, as illustrated in Fig.~\ref{fig:networkslice_lcm_timeline}. In this scenario, different actions take place at different time instants. For instance, at time instants $T_0$, $T_1$, and $T_2$, the NSIs for security cameras, organizers, and participants are created, respectively. Subsequently, at time instants $T_3$ and $T_4$, the NSIs for organizers and participants are updated, respectively. At $T_5$, the organizers’ NSI is further upgraded by changing the use-case category from video chat to push-to-talk. At the end of the event, as participants and organizers leave the campus, their NSIs are terminated in reverse order of their creation. The entire scenario is executed 20 times to assess reproducibility and reduce the impact of experimental variability. 

\begin{figure*}[!t]
    \centering
    % ---------- First row: right-aligned legends PDF ----------
    \hfill
    \begin{subfigure}[t]{0.85\textwidth}
        \raggedleft
        \includegraphics[width=\textwidth]{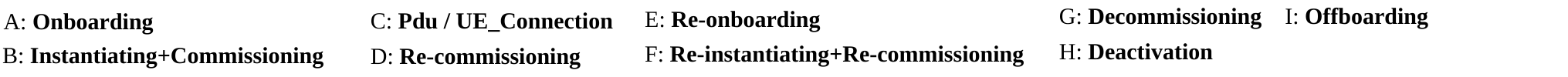}
        \label{fig:lcm_legends}
    \end{subfigure}

    \vspace{-4mm}

    % ---------- Second row: lifecycle management figures ----------
    \begin{subfigure}[t]{0.5\textwidth}
        \centering
        \includegraphics[width=\textwidth]{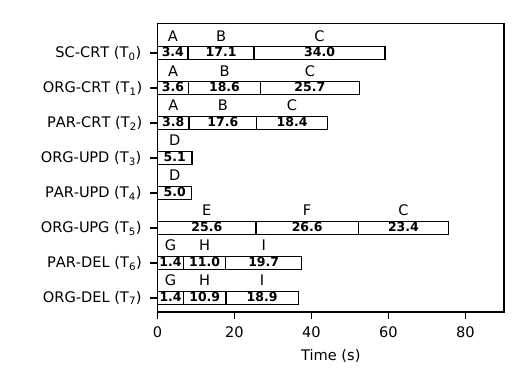}
        \caption{Average NSI lifecycle management execution times.}
        \label{fig:networkslice_lcm_result}
    \end{subfigure}
    \hfill
    \begin{subfigure}[t]{0.49\textwidth}
        \centering
        \includegraphics[width=\textwidth]{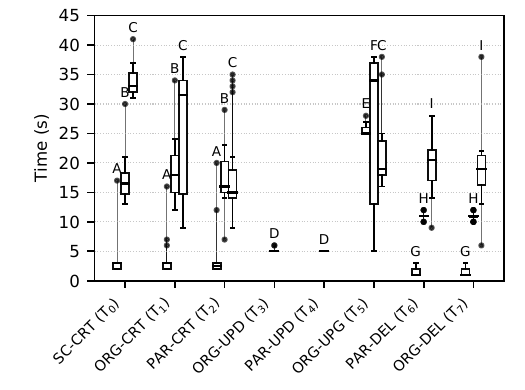}
        \caption{Distribution of NSI lifecycle management execution times across 20 runs.}
        \label{fig:box_chart_lcm_result}
    \end{subfigure}

    \caption{Campus Event NSI lifecycle management results.}
    \label{fig:lcm_results_combined}
\end{figure*}

Before presenting the results, we note that the NSI lifecycle management operations defined in 3GPP TS 28.530~\cite{ts28530}---namely creation, modification (update or upgrade), and termination---are abstractly specified. Therefore, we decompose these operations into finer-grained lifecycle management metrics of NSIs to enable a more precise measurement of their execution time. Furthermore, to measure these metrics in a cloud-native environment, we propose measurement models, consistent with their definitions, that derive them primarily from Kubernetes-reported timestamps, complemented by runtime events at the UE and QoS-enforcement plane. Table~\ref{tab:cloudnative_lcm_metrics} presents these metrics along with their corresponding measurement models, and Table~\ref{tab:notation-lcm_results} defines the notation used in these models. To the best of our knowledge, these cloud-native lifecycle management metrics for NSIs are presented for the first time in this paper. These metrics can be used to evaluate the lifecycle management of similar cloud-native NSI orchestrators. Using the lifecycle management metrics as legend items, Fig.~\ref{fig:networkslice_lcm_result} reports the average results across the 20 runs of the scenario in Fig.~\ref{fig:networkslice_lcm_timeline}. The corresponding distributions and summary statistics are shown as box plots in Fig.~\ref{fig:box_chart_lcm_result}.

%%%%%%%%%%%%%%%%%%%%%%%%%%%%%%%
%%%%%%%%%%%%%%%%%%%%%%%%%%%%%%%
%%%%%%%%%%%%%%%%%%%%%%%%%%%%%%%
%%%%%%%%%%%%%%%%%%%%%%%%%%%%%%%

%%%%%%%%%%%%%%%%%%%%%%%%%%%%%%%%%%%%%%%%%%
%%%%%%%%%%%%%%%%%%%%%%%%%%%%%%%%%%%%%%%%%%
%%%%%%%%%%%%%%%%%%%%%%%%%%%
%%%%%%%%%%%%%%%%%%%%%%%%%%%

As discussed in Section~\ref{Sec:design_overview}, each Deployment represents a single NF in the Kubernetes cluster. To be precise, creating, updating, or deleting a Deployment is the outcome of the orchestration logic, while those operations are realized in the corresponding Pods, managed transitively through the ReplicaSet. Therefore, when NSIs for the corresponding Service Profiles are created by mapping them to NFs, the \textbf{creationTimestamp} of a Deployment indicates the instantiation of an NF, while the \textbf{lastUpdatedTime} or \textbf{deletionTimestamp} of a Deployment indicates a slicing-related update or deletion applied to the corresponding NF. Accordingly, the readiness or deletion timestamp of the corresponding Pods marks the realization of these operations.

%%%%%%%%%%%%%%%%%%%%%%%%%%%%%%%%%%%%%%%%%%%%%%%%%%%%%%%%%%%%%%%%%%
%%%%%% Description Lifecycle management metrics of NSIs %%%%%%%%%%
%%%%%%%%%%%%%%%%%%%%%%%%%%%%%%%%%%%%%%%%%%%%%%%%%%%%%%%%%%%%%%%%%%
\begin{table}[!t]
\caption{Lifecycle management metrics of NSIs for cloud-native environments.}
\label{tab:cloudnative_lcm_metrics}
\centering
\begin{tabular}{|@{\hspace{3pt}}p{2.2cm}|p{5.6cm}|}
\hline
\textbf{Term} & \textbf{Description / Formula} \\
\hline

Onboarding &
Mapping an NSI to NFs.
\newline\newline
\scalebox{0.92}{$
T_{\mathrm{onboard}}(SP_i)
= \left|T_{SP_i}^{\mathrm{create}}
- T_{\mathrm{D}}^{\max}(NSI_i)\right|
$} \\ \hline

Instantiating+\allowbreak commissioning &
Instantiating NFs or reconfiguring them to establish an NSI
\newline\newline
\scalebox{0.92}{$
T_{\mathrm{inst}}(SP_i)
= \left|T_{\mathrm{D}}^{\max}(NSI_i)
- T_{\mathrm{Pod}}^{\max}(NSI_i)\right|
$} \\ \hline

PDU / UE Connection &
Validating NSI by establishing a PDU session.
\newline\newline
\scalebox{0.92}{$
T_{\mathrm{PDU}}(SP_i)
= \left|T_{\mathrm{Pod}}^{\max}(NSI_i)
- T_{\mathrm{UE}}^{\mathrm{PDU}}(NSI_i)\right|
$} \\ \hline

Re-commissioning &
Reconfiguring NFs.
\newline\newline
\scalebox{0.92}{$
T_{\mathrm{re\text{-}comm}}(SP_i)
= \left|T_{SP_i}^{\mathrm{modify}}
- T_{\mathrm{QoS}}^{\mathrm{reinf}}(SP_i)\right|
$} \\ \hline

Re-onboarding &
Remapping an NSI to NFs.
\newline\newline
\scalebox{0.92}{$
T_{\mathrm{re\text{-}onboard}}(SP_i)
= \left|T_{SP_i}^{\mathrm{modify}}
- T_{\mathrm{D}}^{\max}(NSI_i)\right|
$} \\ \hline

Re-instantiating+re-commissioning &
Reinstantiating or reconfiguring NFs after re-onboarding.
\newline\newline
\scalebox{0.92}{$
T_{\mathrm{re\text{-}inst}}(SP_i)
= \left|T_{\mathrm{D}}^{\max}(NSI_i)
- T_{\mathrm{Pod}}^{\max}(NSI_i)\right|
$} \\ \hline

Decommissioning &
Removing NSI from serving NFs.
\newline\newline
\scalebox{0.92}{$
T_{\mathrm{decomm}}(SP_i)
= \left|T_{SP_i}^{\mathrm{delete}}
- T_{\mathrm{D}}^{\max}(NSI_i)\right|
$} \\ \hline

Deactivation &
Disconnecting UEs and disabling an NSI.
\newline\newline
\scalebox{0.92}{$
T_{\mathrm{deact}}(SP_i)
= \left|T_{\mathrm{D}}^{\max}(NSI_i)
- T_{\mathrm{UE}}^{\mathrm{discon}}(NSI_i)\right|
$} \\ \hline

Offboarding &
Scaling down or reconfiguring NFs to delete an NSI.
\newline\newline
\scalebox{0.92}{$
T_{\mathrm{offboard}}(SP_i)
= \left|T_{\mathrm{D}}^{\max}(NSI_i)
- T_{\mathrm{Pod}}^{\mathrm{purge}}(NSI_i)\right|
$} \\ \hline

\end{tabular}
\end{table}
%%%%%%%%%%%%%%%%%%%%%%%%%%%%%%%%%%%%%%%%%%%%%
%%%%%%%%%%%%%%%% End %%%%%%%%%%%%%%%%%%%%%%%%
%%%%%%%%%%%%%%%%%%%%%%%%%%%%%%%%%%%%%%%%%%%%%

%%%%%%%%%%%%%%%%%%%%%%%%%%%%%%%%%%%%%%%%%%%%%
%%%% Notation used for legend items formula%%
%%%%%%%%%%%%%%%%%%%%%%%%%%%%%%%%%%%%%%%%%%%%%
\begin{table}[!t]
\caption{Notation Used in Table~\ref{tab:cloudnative_lcm_metrics} }
\label{tab:notation-lcm_results}
\centering
\renewcommand{\arraystretch}{1.5}
\begin{tabular}{l p{0.65\columnwidth}}
\hline
\textbf{Symbol} & \textbf{Description} \\
\hline

$SP_i$ &
$i$-th Service Profile \\

$NSI_i$ &
$NSI_i$ corresponding to $SP_i$ \\

$\mathcal{N}(NSI_i)$ &
Set of NFs serving $NSI_i$ \\

$NF_j$ &
$j$-th NF, where $NF_j \in \mathcal{N}(NSI_i)$ \\

$\displaystyle \max_{\mathcal{N}(NSI_i)}(\cdot)$ &
Maximum operator taken over all $NF_j \in \mathcal{N}(NSI_i)$ \\

$T_{SP_i}^{\mathrm{create}}$ &
creationTimestamp of $SP_i$ \\

$T_{SP_i}^{\mathrm{modify}}$ &
lastModifiedTime of $SP_i$ \\

$T_{SP_i}^{\mathrm{delete}}$ &
deletionTimestamp of $SP_i$ \\

$T_{\mathrm{D},j}^{\mathrm{create}}$ &
creationTimestamp of the Deployment corresponding to $NF_j$ \\

$T_{\mathrm{D},j}^{\mathrm{update}}$ &
lastUpdatedTime of the Deployment corresponding to $NF_j$ \\

$T_{\mathrm{D},j}^{\max}$ &
$\max\!\left(T_{\mathrm{D},j}^{\mathrm{create}},\,
T_{\mathrm{D},j}^{\mathrm{update}}\right)$ \\

$T_{\mathrm{D}}^{\max}(NSI_i)$ &
$\max_{\mathcal{N}(NSI_i)}\!\left(T_{\mathrm{D},j}^{\max}\right)$ \\

$T_{\mathrm{Pod},j}^{\mathrm{ready}}$ &
lastTransitionTime for the Pod corresponding to $NF_j$ \\

$T_{\mathrm{Pod},j}^{\mathrm{delete}}$ &
deletionTimestamp for the Pod corresponding to $NF_j$ \\

$T_{\mathrm{Pod}}^{\max}(NSI_i)$ &
$\max_{\mathcal{N}(NSI_i)}\!\left(T_{\mathrm{Pod},j}^{\mathrm{ready}}\right)$ \\

$T_{\mathrm{Pod}}^{\mathrm{purge}}(NSI_i)$ &
$\max_{\mathcal{N}(NSI_i)}\!\left(
T_{\mathrm{Pod},j}^{\mathrm{ready}},\,
T_{\mathrm{Pod},j}^{\mathrm{delete}}
\right)$ \\

$T_{\mathrm{UE}}^{\mathrm{PDU}}(NSI_i)$ &
Time at which a UE establishes a PDU session for $NSI_i$ \\

$T_{\mathrm{UE}}^{\mathrm{discon}}(NSI_i)$ &
Time at which a UE is disconnected from $NSI_i$ \\

$T_{\mathrm{QoS}}^{\mathrm{reinf}}(SP_i)$ &
Time at which QoS is re-enforced for $SP_i$ \\

\hline
\end{tabular}
\end{table}
%%%%%%%%%%%%%%%%%%%%%%%%%%%%%%%%%%%%%%%%%%%%%
%%%%%%%%%%%%%%%% End %%%%%%%%%%%%%%%%%%%%%%%%
%%%%%%%%%%%%%%%%%%%%%%%%%%%%%%%%%%%%%%%%%%%%%

In Fig.~\ref{fig:networkslice_lcm_result}, the \textbf{PDU/UE Connection} metric reflects the UE-driven time to (re)establish a PDU session for NSI verification and is not controlled by METIS. The \textbf{(re)instantiating+(re)commissioning} metric captures the time required by the Kubernetes cluster and NFMF agents to deploy and initialize the Pods hosting NFs; once this phase completes, the corresponding NSI can be considered created. As explained in Sub-subsection~\ref{subsubsection:nsi_modification}, upgrading an NSI requires terminating the established NSI and recreating it with the new features. The re-onboarding phase of an NSI upgrade therefore combines termination and creation operations, making the re-onboarding time longer than the onboarding time in an NSI creation procedure. The same applies to the \textbf{re-instantiation+re-commissioning} phase during upgrades: although its measurement model mirrors that of the \textbf{instantiation+commissioning} phase during NSI creation, it reports the time taken for the new NSI to be revived. By considering the above explanation, the fine-grained NSI lifecycle management operations are defined as follows, and Table~\ref{tab:ns_lifecycle_management_operations_latency} reports the measured execution time using the results of Fig.~\ref{fig:networkslice_lcm_result}.

\begin{itemize}
  \item \textbf{Creation time}: defined as the sum of the onboarding time and the instantiating+commissioning time.
  \item \textbf{Update time}: defined as the re-commissioning time.
  \item \textbf{Upgrade time}: defined as the sum of the re-onboarding time and the re-instantiating+re-commissioning time.
  \item \textbf{Deletion time}: defined as the sum of the decommissioning, deactivation, and offboarding times.
\end{itemize}

The results demonstrate the ability of METIS in dynamically performing NSI lifecycle management operations, thereby enabling NSIs to adapt to the evolving behavior of applications and services over time. For instance, the total time required to create an NSI is approximately 22.4~s, whereas updating, upgrading, and deleting an NSI take approximately 5.1~s, 52.2~s, and 32.1~s, respectively. Notably, the measured upgrade latency ($\approx$ 52.2~s) closely approximates the sum of the creation and deletion latencies ($\approx$ 54.5~s), consistent with the fact that METIS realizes an upgrade as a termination of the existing NSI followed by instantiation of the new one. This correspondence serves as an internal consistency check on the proposed lifecycle measurement models. 

Table~\ref{tab:ns_lifecycle_management_operations_latency} also positions METIS against two cloud-native NSI orchestration baselines, NASP~\cite{nasp} and CLiSO~\cite{cliso}. Two distinctions are important when interpreting this comparison. First, the scale differs: METIS results are measured for three concurrent NSIs, whereas both baselines report a single slice, making the comparison conservative. Second, the scope of orchestration differs. NASP primarily renders and deploys a slice template, reporting only a per-scenario creation time and a single AMF replace-and-redeploy reconfiguration ($\approx$9~s outage); it does not perform a managed update, a true category upgrade, or a measured termination. METIS, in contrast, executes each TS~28.530 operation as a managed, decomposed phase. CLiSO reports creation and deletion for a multi-domain slice but issues an asynchronous \texttt{DELETE} over the Kubernetes REST API, which returns once the request is accepted rather than once the Pods are fully torn down; its $\approx$1--3.5~s deletion therefore reflects deletion \emph{requested}, whereas the METIS deletion time ($\leq$32.1~s) measures decommissioning, deactivation, and offboarding through to complete teardown.

%%%%%%%%%%%%%%%%%%%%%%%%%%%%%%%%%%%%%%%%%%%%%%%%%%%%%%%%
%%Network Slice Lifecycle Management Operation Latency%%
%%%%%%%%%%%%%%%%%%%%%%%%%%%%%%%%%%%%%%%%%%%%%%%%%%%%%%%%

\begin{table}[!t]
\centering
\footnotesize
\setlength{\tabcolsep}{3pt}
\caption{Network Slice Lifecycle Management Operation Latency (s).}
\label{tab:ns_lifecycle_management_operations_latency}
\begin{tabular}{@{}llccc@{}}
\toprule
\textbf{Operation} & \textbf{Phase} & \textbf{METIS} & \textbf{NASP}\,\cite{nasp} & \textbf{CLiSO}\,\cite{cliso} \\
\midrule
\multirow{3}{*}{Creation}
 & Onboarding                & $\leq 3.8$  & -- & -- \\
 & Inst.+Comm.               & $\leq 18.6$ & -- & -- \\
 & \textbf{Total}            & $\leq 22.4$ & $22$--$53^{\dagger}$ & $43.0^{\ddagger}$ \\
\midrule
Update
 & \textbf{Total} (Re-comm.) & $\leq 5.1$  & $\approx 9^{\S}$ & -- \\
\midrule
\multirow{3}{*}{Upgrade}
 & Re-onboarding             & $\leq 25.6$ & -- & -- \\
 & Re-inst.+Re-comm.         & $\leq 26.6$ & -- & -- \\
 & \textbf{Total}            & $\leq 52.2$ & -- & -- \\
\midrule
\multirow{4}{*}{Deletion}
 & Decommissioning           & $\leq 1.4$  & -- & -- \\
 & Deactivation              & $\leq 11.0$ & -- & -- \\
 & Offboarding               & $\leq 19.7$ & -- & -- \\
 & \textbf{Total}            & $\leq 32.1$ & -- & $3.5^{\ddagger}$ \\
\bottomrule
\end{tabular}
\vspace{2pt}

\parbox{\columnwidth}{\raggedright\footnotesize
$^{\dagger}$ Scenario-dependent (Shared $\approx$22, mMTC $\approx$42,
non-3GPP $\approx$50, URLLC $\approx$53).\\
$^{\ddagger}$ Table~4 in~\cite{cliso}.\\
$^{\S}$ Similar to NSI update: AMF replace-and-redeploy reconfiguration outage.}
\end{table}

\textbf{\textit{Takeaway.}} The results show that declarative, reconciliation-driven orchestration enables dynamic NSI lifecycle management: not only Day-0/1 design and instantiation but also Day-2 adaptation and termination, allowing NSIs to evolve with changing network conditions and application requirements. With three concurrent NSIs, METIS completes creation, update, upgrade, and deletion in at most 22.4~s, 5.1~s, 52.2~s, and 32.1~s, respectively.

\subsection{Joint RAN and Core Network Slicing with QoS Enforcement} \label{subsection:ran_and_cn_slicing}

In this subsection, we evaluate METIS in terms of its ability to provide joint RAN and CN slicing with QoS enforcement, as defined in Table~\ref{tab:campus-event-use-case} for the campus event. We conduct an overload test in which each UE, belonging to its corresponding NSI, simultaneously transmits traffic exceeding its declared QoE. The purpose of the overload test is not only to evaluate the isolation of established NSIs but also to assess whether traffic is constrained within the specified QoE limits, thereby ensuring that the SLAs of the NSIs are met. To compare the results of this section, we also conducted two more scenarios: (1) Without RAN and without CN slicing, and (2) Without RAN slicing but with CN slicing. 

%%%%%%%%%%%%%%%%%%%%%%%% Slicing Scenarios table %%%%%%%%%%
\begin{table*}[!t]
\centering
\caption{Average throughput and SLA ratio for DL/UL (TCP/UDP) under overload across slicing configurations.}
\label{tab:slicing_scenarios}
\renewcommand{\arraystretch}{1.1}
\setlength{\tabcolsep}{5pt}

\begin{tabular}{llcccccc@{\hspace{6pt}}cccccc}
\toprule
\multirow{4}{*}{\textbf{Scenario}} 
& \multirow{4}{*}{\textbf{Slice}} 
& \multicolumn{5}{c}{\textbf{Downlink (DL)}} 
& \multicolumn{5}{c}{\textbf{Uplink (UL)}} \\
\cmidrule(lr){3-7} \cmidrule(lr){9-13}

&
& \textbf{Target} 
& \multicolumn{4}{c}{\textbf{Measured}} 
& 
& \textbf{Target} 
& \multicolumn{4}{c}{\textbf{Measured}} \\
\cmidrule(lr){4-7} \cmidrule(lr){10-13}

&
&
& \multicolumn{2}{c}{\textbf{TCP}}
& \multicolumn{2}{c}{\textbf{UDP}}
&
&
& \multicolumn{2}{c}{\textbf{TCP}}
& \multicolumn{2}{c}{\textbf{UDP}} \\
\cmidrule(lr){4-5} \cmidrule(lr){6-7} \cmidrule(lr){10-11} \cmidrule(lr){12-13}

&
& \textbf{Avg.}
& \textbf{Avg.} & \textbf{SLA}
& \textbf{Avg.} & \textbf{SLA}
&
& \textbf{Avg.}
& \textbf{Avg.} & \textbf{SLA}
& \textbf{Avg.} & \textbf{SLA} \\

&
& (Mbps)
& (Mbps) & (\%)
& (Mbps) & (\%)
&
& (Mbps)
& (Mbps) & (\%)
& (Mbps) & (\%) \\
\midrule

\multirow{3}{*}{No slicing}
& Security Cameras
& 10
& 54.15 & 0.17 \xmark
& 54.57 & 0.17 \xmark
&
& 10
& 10.62 & 71.33 \xmark
& 10.79 & 67.67 \xmark \\
& Organizers
& 50
& 53.37 & 98.12 \cmark
& 53.08 & 89.50 \cmark
&
& 8
& 11.00 & 9.17 \xmark
& 10.97 & 15.33 \xmark \\
& Participants
& 40
& 54.24 & 0.68 \xmark
& 53.99 & 3.50 \xmark
&
& 3
& 10.58 & 0.00 \xmark
& 11.02 & 0.00 \xmark \\
\midrule

\multirow{3}{*}{CN only}
& Security Cameras
& 10
& 9.94 & 84.14 \cmark
& 10.12 & 99.33 \cmark
&
& 10
& 9.90 & 64.17 \xmark
& 8.31 & 51.50 \xmark \\
& Organizers
& 50
& 47.24 & 80.67 \cmark
& 50.30 & 99.33 \cmark
&
& 8
& 7.93 & 67.67 \xmark
& 8.05 & 91.17 \cmark \\
& Participants
& 40
& 39.00 & 84.67 \cmark
& 40.22 & 99.67 \cmark
&
& 3
& 3.00 & 71.50 \xmark
& 3.05 & 99.67 \cmark \\
\midrule

\multirow{3}{*}{RAN+CN}
& Security Cameras
& 10
& 9.46 & 94.83 \cmark
& 9.86 & 96.50 \cmark
&
& 10
& 9.61 & 81.50 \cmark
& 9.77 & 88.33 \cmark \\
& Organizers
& 50
& 47.48 & 86.83 \cmark
& 49.03 & 96.67 \cmark
&
& 8
& 7.68 & 84.67 \cmark
& 7.68 & 88.50 \cmark \\
& Participants
& 40
& 38.08 & 86.83 \cmark
& 39.27 & 97.17 \cmark
&
& 3
& 2.73 & 80.67 \cmark
& 2.71 & 81.17 \cmark \\
\bottomrule
\end{tabular}

\vspace{2pt}
\begin{minipage}{\textwidth}
\centering
\footnotesize
\textit{Note:} \cmark\ indicates SLA $\geq$ TCT, while \xmark\ indicates SLA $<$ TCT.
\end{minipage}
\end{table*}
%%%%%%%%%%%% end %%%%%%%%%%%%%%

%%%%%%%%%%%%%%%%%%%%%%%%%%%%%%
%%%%%%%%%%%%%%%%%%%%%%%%%%%%%%
\begin{figure*}[!t]
    \centering

    % ---------- First row: centered legends PDF ----------
    \begin{subfigure}[t]{0.55\textwidth}
        \centering
        \includegraphics[width=0.9\textwidth]{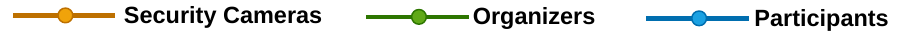}
        \label{fig:placeholder_image}
    \end{subfigure}%

    % ---------- Second row: throughput figures ----------
    \begin{subfigure}[t]{0.24\textwidth}
        \centering
        \includegraphics[width=\textwidth]{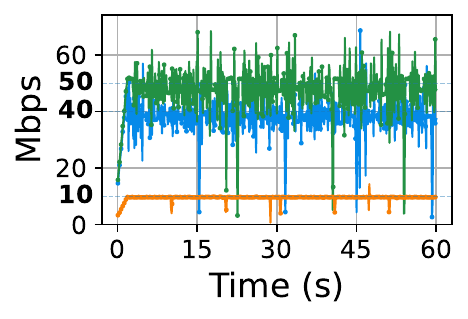}
        \caption{DL TCP}
        \label{fig:dl_tcp}
    \end{subfigure}
    \hfill
    \begin{subfigure}[t]{0.24\textwidth}
        \centering
        \includegraphics[width=\textwidth]{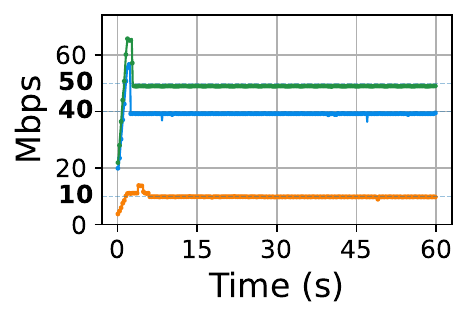}
        \caption{DL UDP}
        \label{fig:dl_udp}
    \end{subfigure}
    \hfill
    \begin{subfigure}[t]{0.24\textwidth}
        \centering
        \includegraphics[width=\textwidth]{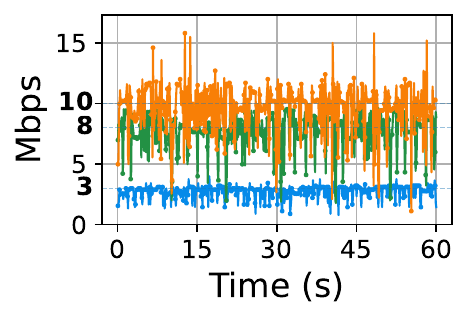}
        \caption{UL TCP}
        \label{fig:ul_tcp}
    \end{subfigure}
    \hfill
    \begin{subfigure}[t]{0.24\textwidth}
        \centering
        \includegraphics[width=\textwidth]{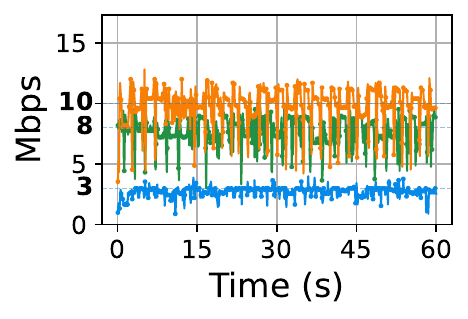}
        \caption{UL UDP}
        \label{fig:ul_udp}
    \end{subfigure}

    \caption{Throughput time-series for DL/UL (TCP/UDP) with joint RAN and CN slicing.}
    \label{fig:throughput_ul_dl_tcp_udp}
\end{figure*}
%%%%%%%%%%%%%%%%%%%%%%%%%%%%%%%%%%%%%%%%%%%%%%
%%%%%%%%%%%%%%%%%%%%%%%%%%%%%%%%%%%%%%%%%%%%%%

Before presenting the results, we describe the SLA ratio (SLA(\%)) metric, defined in \eqref{eq:sla_formula}. This metric assesses whether the requested QoE for an NSI is satisfied during its observation period from the Slice Operator’s perspective, regardless of the QoS enforcement mechanisms applied at the RAN and CN. To this end, starting from the initiation of an NSI, time-series traffic samples are collected, where the function $Traffic(t_n)$ represents the downlink or uplink throughput at the sampling instant $t_n$. Let $t_0, t_1, \dots, t_{N-1}$ denote the discrete sampling instants, and $N$ denotes the total number of captured traffic samples from $t_0$ to $t_{N-1}$ in \eqref{eq:sla_formula}, with a sampling interval of 100 milliseconds. Depending on an NSI, $t_{N-1}$ corresponds to the reporting period declared by the use case owner in its Service Profile. The analysis is performed over time windows of duration $t_{N-1} - t_0$, repeated over the NSI lifetime.

Under the overload test, a traffic sample at $t_n$ is considered valid and assigned the value 1 in \eqref{eq:sla_formula} if it lies within the range defined by the guaranteed traffic and the maximum allowable traffic for the NSI, as specified in Table~\ref{tab:campus-event-use-case}. Therefore, the SLA ratio for an NSI is computed as the ratio of valid samples to the total number of captured samples during the observation period of the NSI. This value must be greater than or equal to the required time compliance threshold $TCT$ specified in Table~\ref{tab:campus-event-use-case}; otherwise, the SLA status is considered not satisfied for the NSI over the considered time window.

\begin{equation}
\label{eq:sla_formula}
\small
\begin{aligned}
\text{SLA}(\%) &=
\frac{
\sum_{n=0}^{N-1}
\begin{cases}
1, & \text{if } \text{GTD} \leq \text{Traffic}(t_n) \leq \text{MAX} \\
0, & \text{otherwise}
\end{cases}
}{N}
\times 100
\end{aligned}
\end{equation}

Therefore, in the overload test, each of the three Quectel modules (UEs) was connected to its corresponding NSI in the Campus Event use case and generated downlink and uplink traffic exceeding its QoE, namely 100~Mbps for downlink and 20~Mbps for uplink. In this context, an observation window with a duration of one minute—during which all three UEs simultaneously transmit traffic—was considered. This yields 600 samples (with a sampling interval of 100 milliseconds) for each of the four traffic conditions: TCP downlink, TCP uplink, UDP downlink, and UDP uplink. This test was repeated for the other two scenarios as well; 1) Without RAN and without CN slicing, and 2) Without RAN slicing but with CN slicing; and Table~\ref{tab:slicing_scenarios} presents the average throughput and SLA ratio measured over the collected samples under the overload test for all three scenarios—No slicing, CN only, and RAN+CN—across the four traffic conditions.

Given that, in the RAN+CN scenario shown in Table~\ref{tab:slicing_scenarios}, the measured SLA ratio metric for all four traffic conditions is greater than or equal to the defined $TCT$ in Table~\ref{tab:campus-event-use-case}, the SLA is satisfied under the overload test. 
In contrast, in the No slicing scenario, the SLA is violated in almost all traffic conditions. This is because there is neither isolation nor QoS enforcement, and network resources are fairly shared among UEs.

In the CN only scenario, isolation and QoS enforcement are applied only at the CN. As a result, downlink traffic is controlled for NSIs based on their QoE, and the measured SLA ratio metric is satisfied for both TCP and UDP downlink traffic. However, since there is neither isolation nor QoS enforcement on the RAN side, RAN resources are fairly shared among UEs across NSIs. Consequently, for uplink traffic, the SLA is violated for TCP across all NSIs, while for UDP, it is violated only for the security cameras' slice and remains satisfied for the other NSIs. These results highlight the importance of joint RAN and CN slicing with QoS enforcement, and show why the isolation among NSIs configured by the Slice Operator is key for satisfying SLAs.

In the following paragraphs, we analyze the behavior of transport protocols—TCP and UDP—under both downlink and uplink traffic conditions. To this end, Fig.~\ref{fig:throughput_ul_dl_tcp_udp} shows the 600 time-series traffic samples for each of the four traffic conditions for the RAN+CN scenario.

Based on the subfigures in Fig.~\ref{fig:throughput_ul_dl_tcp_udp}, several questions may arise. To address some of them, we performed a detailed analysis. The first question is why Fig.~\ref{fig:dl_udp} exhibits an almost flat downlink traffic pattern that closely follows the requested QoE for the NSIs. The main reason is that Linux Traffic Control is configured and applied to the dedicated tunnel network interface (TUN) at the UPFs in the CN for each NSI, based on the requested QoE. This effectively shapes and throttles the downlink traffic in the overload test before it reaches the RAN; thus, the traffic arriving at the RAN is already regulated, resulting in stable and nearly constant throughput. Note that these experiments were conducted under stable channel conditions: the over-the-air interface operated in band n48, no external interference was observed, and the measured RSRP at the UEs was approximately $-60$~dBm.

In contrast, Fig.~\ref{fig:ul_udp} shows that UDP-based uplink traffic does not exhibit the same flat pattern, despite sufficient PRBs being allocated to each NSI in the RAN via the xApp based on its QoE. This is because the configured PRB percentage only determines the average share of radio resources that the MAC scheduler should allocate to each NSI over time. It does not regulate the traffic generation rate at the UE side and does not perform traffic shaping within the RAN. Therefore, it does not enforce a strict PRB reservation or rate limit. 

In the uplink direction, traffic originates from UE-side applications, which inject UDP traffic at a maximum rate of 20~Mbps, and reaches the 5G UE modems, i.e., the Quectel modules, without prior regulation. As a result, the RLC TX buffers of the Quectel modules fill faster than the allocated PRB percentage can drain them. When the MAC scheduler receives Buffer Status Reports (BSRs), it observes only the volume of buffered uplink data awaiting transmission, with no visibility into the application-level target rate of each NSI—e.g., 3~Mbps for participants, 8~Mbps for organizers, or 10~Mbps for security cameras. As a result, the scheduler cannot distinguish a legitimately bursty application from an unregulated constant-rate UDP source with a saturated buffer. 

Regarding the behavior observed in Fig.~\ref{fig:dl_tcp} and Fig.~\ref{fig:ul_tcp}, the main reason behind the fluctuations around the requested average traffic for all three NSIs is the congestion control mechanism applied by TCP.

{\textbf{\textit{Takeaway.}} The three slicing configurations show that joint RAN and CN slicing is necessary, not merely beneficial. CN-side shaping can satisfy the downlink SLA by regulating traffic before it reaches the RAN, but it cannot satisfy the uplink SLA because UE-originated traffic reaches the RAN unregulated and the MAC scheduler cannot impose per-NSI rates from BSR reports alone. Thus, only the RAN+CN configuration meets the TCT across all four traffic conditions. More broadly, NSI resource allocation alone cannot guarantee the required QoS; SLA-aware traffic shaping must also be enforced across the UE, RAN~\cite{tcran}, and CN.

% \textcolor{red}{You need to confirm your claim of joint RAN and Core: add UL and DL case and the need for joint slicing.}

%%%%%%%%%%%%%%%%%%%%%%%
%%%%%%%%%%%%%%%%%%%%%%%
\begin{figure*}[!t]
  \centering
   \label{fig:scalability_results}
  \begin{subfigure}[t]{0.32\textwidth}
    \centering
    \includegraphics[width=\linewidth]{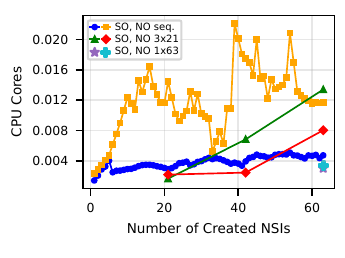}
    \caption{SO and NO CPU usage}
    \label{fig:cpu_usage_so_no}
  \end{subfigure}
  \hfill
  \begin{subfigure}[t]{0.32\textwidth}
    \centering
    \includegraphics[width=\linewidth]{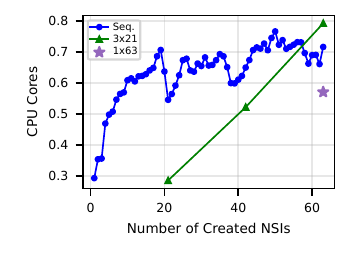}
    \caption{Kubernetes API server CPU usage}
    \label{fig:kubernetes_apiserver_cpu}
  \end{subfigure}
  \hfill
  \begin{subfigure}[t]{0.32\textwidth}
    \centering
    \includegraphics[width=\linewidth]{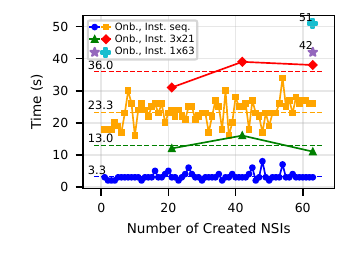}
    \caption{Onboarding and instantiation+commissioning times}
    \label{fig:onboarding_instantiating}
  \end{subfigure}

  \caption{Scalability results in terms of CPU Utilization and NSI creation time.}

\end{figure*}
%%%%%%%%%%%%%%%%%%%%%%%
%%%%%%%%%%%%%%%%%%%%%%%

\subsection{Scalability} \label{subsection:scalability}

To evaluate the scalability of METIS under increasing orchestration complexity in distributed environments, a multi-region deployment scenario is considered, consisting of three regions with three zones each, i.e., nine geographically distributed zones in total. In this setup, 63 NSIs are deployed under three different scenarios: (1) sequential deployment of 63 NSIs, (2) parallel deployment in three batches of 21 NSIs, and (3) parallel deployment of all 63 NSIs in a single batch. The choice of 63 NSIs is constrained by the available infrastructure, namely the nine zones, and by the limitations of the adopted 5G CN, Open5GS, which supports up to seven network slices per SMF instance. For each scenario, we measure the CPU utilization of the Slice Operator (SO), the Network Operator (NO), and the Kubernetes API server, as well as the corresponding NSI creation time. 

The CPU utilization results for both the SO and NO across the three scenarios are shown in Fig.~\ref{fig:cpu_usage_so_no}, while the corresponding results for the Kubernetes API server are presented in Fig.~\ref{fig:kubernetes_apiserver_cpu}. A fine-grained breakdown of NSI creation, including onboarding and instantiation+commissioning, is shown in Fig.~\ref{fig:onboarding_instantiating} for the same scenarios.

As shown in Fig.~\ref{fig:cpu_usage_so_no}, in the sequential deployment scenario, the CPU utilization of both the SO and NO gradually increases as the number of created NSIs grows. In contrast, when the NSIs are deployed in parallel using three batches of 21 NSIs, the CPU utilization shows an approximately linear increase across the batches. Interestingly, the CPU load observed during the deployment of the third batch of 21 NSIs is higher than that observed during the simultaneous deployment of all 63 NSIs. This behavior can be explained by the dependency between newly requested NSIs and already deployed NSIs. In the sequential scenario, each step processes one new NSI request while the number of active NSIs accumulates over time. In contrast, when all 63 NSIs are deployed in a single batch, all requests are processed simultaneously, with no previously deployed NSIs in the system. This indicates that the CPU utilization of the SO and NO depends not only on the number of newly arriving NSIs processed in parallel, but also on the number of NSIs already deployed. This interpretation is supported by the results shown in Fig.~\ref{fig:kubernetes_apiserver_cpu}, where the CPU utilization of the Kubernetes API server is higher during the deployment of the third 21-NSI batch compared to the other two scenarios. Note that the Kubernetes API server does not exclusively serve the SO and NO: its measured utilization includes the cluster's baseline control-plane load ($\approx$0.3 CPU cores before any NSI is created), so the METIS-attributable portion is the increment above this baseline, a cost inherent to any operator-based orchestrator interacting with the Kubernetes API.

Furthermore, Fig.~\ref{fig:onboarding_instantiating} shows how the deployment scenario affects NSI creation time: the mean creation time is 26.6~s for sequential deployment, 49.0~s for the three 21-NSI batches, and 93~s for the single 63-NSI batch. The sequential result is directly comparable to the creation time reported in Table~\ref{tab:ns_lifecycle_management_operations_latency}: sequentially creating 3 NSIs in the campus event takes up to 22.4~s, whereas sequentially creating 63 NSIs here yields a mean of 26.6~s---a difference of roughly 4~s. Decomposing this difference shows that the onboarding times are nearly identical (within 0.5~s), while instantiating+commissioning accounts for approximately 4.7~s of additional time. The reason is scale: the campus event deploys 18 O-RAN/5G NFs (Fig.~\ref{fig:campus_event_deploy}), whereas the scalability test deploys 62 O-RAN/5G NFs across nine zones, increasing the load on cluster nodes and the NFMF agents that resolve inter-NF dependencies in a more complex topology. Finally, Figs.~\ref{fig:cpu_usage_so_no} and~\ref{fig:onboarding_instantiating} together reveal a trade-off: deploying all 63 NSIs in a single batch yields the lowest CPU utilization among the three scenarios but the longest creation time (93~s), whereas sequential deployment inverts both.
   
\textbf{\textit{Takeaway.}} METIS orchestrates 63 NSIs across nine geographically distributed zones with modest control-plane overhead ($\leq$0.03 CPU cores for the SO and NO combined). The deployment strategy trades CPU utilization against creation time---a single 63-NSI batch minimizes the former while sequential deployment minimizes the latter---and, notably, control-plane load grows with the number of already-deployed NSIs, not merely with the arrival rate of new requests.

\subsection{Failure Recovery} \label{subsection:failure_recovery}

To evaluate how METIS recovers NSIs from failures without human intervention, we use Chaos Mesh~\cite{chaosmesh} to inject failures at the O-RAN NF, CN NF, NS, and NSI levels by removing the corresponding resources from the Kubernetes cluster. The reconciliation loops of METIS then observe the actual state of each resource, compare it against the desired state, and act to recover the removed resources.

To perform this evaluation, an NSI is created by applying the corresponding Service Profile, and Chaos Mesh is then used to inject the failures. For each injected failure, Table~\ref{tab:recovery} reports the time for the reconciliation loops to observe the failure ($\leq 0.07$~s), to compare the desired state of the resource against its actual state and act to make them converge ($\leq 0.09$~s), and for the resources to fully recover ($\leq 18.6$~s). Notably, because the orchestration logic is distributed across the reconciliation loops, the loop at the failed level detects and repairs the failure without involving the higher-level loops: when the O-RAN NF or CN NF is deleted, the Network Operator observes and handles the failure, whereas the Slice Operator handles the deletion of the NS and NSI. As shown in the Recovery column of Table~\ref{tab:recovery}, the recovery time for the O-RAN NF (O-gNB) is $\approx$1.5~s longer than that for the CN NF (PCF), while the recovery time for the NSI is around 1~s longer than that for the NS. The former difference arises because the NFMF agent of the O-gNB must resolve two dependencies---the IP addresses of the target AMF and Near-RT RIC---and additionally issue an I/O request to access the USRP serving as the RU. In contrast, the NFMF agent of the PCF resolves only the IP address of the NRF---the Service-Based Architecture allows other NFs, such as the UDM and BSF, to be discovered through it---and requires no I/O request. Regarding the NS and NSI, recovering the NS requires all deleted O-RAN/5G NFs to be re-instantiated and re-commissioned before the NS becomes operational, whereas recovering the NSI requires one additional reconciliation step beyond NS recovery.

\begin{table}[!t]
\centering
\caption{Failure Recovery Times of METIS Across Four Failure Levels (seconds).}
\label{tab:recovery}
\begin{tabular}{lcccc}
\toprule
\textbf{Failure} & \textbf{Observe} & \textbf{Compare/Act} & \textbf{Recovery} & \textbf{Total} \\
\midrule
O-RAN NF  & 0.065 & 0.075 & 7.910 & 8.050 \\
CN NF     & 0.061 & 0.089 & 6.456 & 6.606 \\
NS        & 0.062 & 0.045 & 17.551 & 17.658 \\
NSI       & 0.058 & 0.025 & 18.554 & 18.637 \\
\bottomrule
\end{tabular}
\end{table}

\textbf{\textit{Takeaway.}} METIS recovers NSIs from failures at all four levels autonomously, detecting each failure in under 0.07~s and fully restoring the affected resources in under 19~s. Because orchestration logic is distributed across the cascaded loops, each failure is handled at its own level without escalating to higher loops---localizing recovery and confirming that resilience is a structural property of the reconciliation design rather than a dedicated recovery mechanism.

\section{Conclusion \& Future Work} \label{Sec:conclusion}
This paper presented METIS, a declarative, application-driven Slice Orchestrator that manages NSIs as first-class declarative resources through cascaded reconciliation loops. By extending the modern, declarative orchestration paradigm---previously limited to the scope of NFs and NSs---to the full NSI lifecycle, METIS bridges a critical gap between existing cloud-native orchestration frameworks and the operational demands of 5G-Advanced and future 6G networks. At its core, METIS introduces an application-centric design in which vertical industries express the network requirements of their applications through Service Profiles, through which 3GPP-aligned Slice Profiles are automatically derived. This eliminates reliance on static, preconfigured slice templates and enables adaptive Day-0, Day-1, and Day-2 operations---all realized through a multi-level operator design in a declarative, idempotent, and automated manner.

A central finding of this work is that joint RAN and CN slicing is not merely beneficial but necessary for SLA satisfaction. Under concurrent overload conditions, CN-only slicing fails to control uplink traffic in the absence of RAN-level QoS enforcement, while METIS, coordinating traffic shaping at the UPF with dynamic PRB allocation via the SLA xApp, successfully satisfies the defined SLA for all NSIs across all traffic conditions. This result underscores a fundamental asymmetry in end-to-end slice control: downlink traffic can be shaped at the CN before reaching the RAN, whereas uplink traffic originates at the UE and reaches the RAN unregulated, making RAN-side enforcement indispensable. More broadly, the results confirm that resource allocation alone is insufficient; traffic shaping must be enforced across all three elements of an end-to-end NSI---the UE, the RAN, and the CN.

The lifecycle management evaluation shows the agility of METIS, with NSI creation completing up to 22.4 s and updates requiring up to 5.1 s, enabling NSIs to adapt dynamically to evolving application and service requirements. The scalability evaluation further shows that METIS orchestrates 63 NSIs across a nine-zone, multi-region deployment while consuming well below 0.03 CPU cores, indicating that its cascaded reconciliation loops impose modest control-plane overhead as the number of NSIs grows. The failure recovery results additionally show that METIS observes NSI failures across four levels and fully recovers them in under 19 s. 

Future work will focus on integrating AI/ML- and Digital-Twin-based mechanisms into the SO-Validator for admission control and resource allocation, extending the application-centric service modeling to capture richer traffic behavior beyond rate-based QoE parameters, and evaluating METIS under dynamic NSI workloads in fully disaggregated, multi-vendor O-RAN and CN deployments.

% if have a single appendix:
%\appendix[Proof of the Zonklar Equations]
% or
%\appendix  % for no appendix heading
% do not use \section anymore after \appendix, only \section*
% is possibly needed

% use appendices with more than one appendix
% then use \section to start each appendix
% you must declare a \section before using any
% \subsection or using \label (\appendices by itself
% starts a section numbered zero.)
%

% \appendices
% \section{Proof of the First Zonklar Equation}
% Appendix one text goes here.

% % you can choose not to have a title for an appendix
% % if you want by leaving the argument blank
% \section{}
% Appendix two text goes here.

% use section* for acknowledgment
\section*{Acknowledgment}
This work is supported by the European Commission as part the Horizon Europe 6Green and 6GCloud Projects under Grant 101096925 and 101139073 as well as the  Intention-6G project under the France 2030 programme and co funded by Bpifrance.
The authors would like to thank the BubbleRAN team for their support. They also thank Prof. Roberto Bruschi and Dr. Chiara Lombardo of the University of Genova for their valuable comments during the 6Green project collaboration.
\nocite{*}
\bibliographystyle{IEEEtran}
\bibliography{references}

@article{cliso,
  author  = {S. Arora and A. Ksentini and C. Bonnet},
  title   = {Cloud Native Lightweight Slice Orchestration ({CLiSO}) Framework},
  journal = {Computer Communications},
  volume  = {213},
  pages   = {1--12},
  year    = {2024},
  doi     = {10.1016/j.comcom.2023.10.010}
}

@article{free5gmano,
  author  = {Wei-Cheng Chang and Fuchun Joseph Lin},
  title   = {Coordinated Management of {5G} Core Slices by {MANO} and {OSS/BSS}},
  journal = {Journal of Computer and Communications},
  volume  = {9},
  number  = {6},
  pages   = {52--72},
  year    = {2021},
  doi     = {10.4236/jcc.2021.96004}
}

@article{tcran,
  author  = {Irazabal, Mikel and Nikaein, Navid},
  title   = {TC-RAN: A Programmable Traffic Control Service Model for 5G/6G SD-RAN},
  journal = {IEEE Journal on Selected Areas in Communications},
  volume  = {42},
  number  = {2},
  pages   = {406--419},
  year    = {2024},
  doi     = {10.1109/JSAC.2023.3336162}
}

@article{athena,
  author  = {Mohammadi, Alireza and Nikaein, Navid},
  title   = {Athena: An Intelligent Multi-x Cloud Native Network Operator},
  journal = {IEEE Journal on Selected Areas in Communications},
  volume  = {42},
  number  = {2},
  pages   = {460--472},
  year    = {2024},
  doi     = {10.1109/JSAC.2023.3336172}
}

@article{nasp,
  author  = {Grings, Felipe Hauschild and Bruno, Gustavo Zanatta and Prade, Lucio Rene and Brito, Jos\'{e} Marcos C\^{a}mara and Both, Cristiano Bonato},
  title   = {{NASP}: Network Slice as a Service Platform for {5G} Networks},
  journal = {Journal of Network and Computer Applications},
  volume  = {250},
  pages   = {104479},
  year    = {2026},
  doi     = {10.1016/j.jnca.2026.104479}
}

@inproceedings{oranslice,
  author    = {Cheng, Hai and D'Oro, Salvatore and Gangula, Rajeev and Velumani, Sakthivel and Villa, Davide and Bonati, Leonardo and Polese, Michele and Melodia, Tommaso and Arrobo, Gabriel and Maciocco, Christian},
  title     = {{ORANSlice}: An Open Source {5G} Network Slicing Platform for {O-RAN}},
  booktitle = {Proc. 30th Annu. Int. Conf. Mobile Computing and Networking (ACM MobiCom)},
  pages     = {2297--2302},
  year      = {2024},
  address   = {Washington D.C., USA},
  doi       = {10.1145/3636534.3701544}
}

@inproceedings{flexric,
  author    = {Robert Schmidt and Mikel Irazabal and Navid Nikaein},
  title     = {{FlexRIC}: An {SDK} for Next-Generation {SD-RANs}},
  booktitle = {Proc. 17th Int. Conf. Emerging Networking EXperiments and Technologies (CoNEXT)},
  pages     = {411--425},
  year      = {2021},
  doi       = {10.1145/3485983.3494870}
}

@book{kubernetes_operators_book,
  title     = {Kubernetes Operators: Automating the Container Orchestration Platform},
  author    = {Dobies, Jason and Wood, Joshua},
  publisher = {O'Reilly Media},
  address   = {Sebastopol, CA, USA},
  year      = {2020},
  isbn      = {978-1-4920-4803-9}
}

@book{kubernetes_up_running_book,
  title     = {Kubernetes: Up and Running},
  author    = {Hightower, Kelsey and Burns, Brendan and Beda, Joe},
  publisher = {O'Reilly Media},
  address   = {Sebastopol, CA, USA},
  year      = {2017},
  isbn      = {978-1-4919-3566-8}
}

@techreport{ts23501,
  author      = {{3GPP}},
  title       = {System Architecture for the {5G} System ({5GS})},
  institution = {3rd Generation Partnership Project (3GPP)},
  number      = {TS 23.501},
  year        = {2023},
  note        = {Release 17}
}

@techreport{ts23502,
  author      = {{3GPP}},
  title       = {Procedures for the {5G} System ({5GS})},
  institution = {3rd Generation Partnership Project (3GPP)},
  number      = {TS 23.502},
  year        = {2024},
  note        = {Release 17}
}

@techreport{ts23003,
  author      = {{3GPP}},
  title       = {Numbering, Addressing and Identification},
  institution = {3rd Generation Partnership Project (3GPP)},
  number      = {TS 23.003},
  year        = {2023},
  note        = {Release 17}
}

@techreport{ts28530,
  author      = {{3GPP}},
  title       = {Management and Orchestration; Concepts, Use Cases and Requirements},
  institution = {3rd Generation Partnership Project (3GPP)},
  number      = {TS 28.530},
  year        = {2024},
  note        = {Release 17}
}

@techreport{ts28531,
  author      = {{3GPP}},
  title       = {Management and Orchestration; Provisioning},
  institution = {3rd Generation Partnership Project (3GPP)},
  number      = {TS 28.531},
  year        = {2023},
  note        = {Release 17}
}

@techreport{ts28533,
  author      = {{3GPP}},
  title       = {Management and Orchestration; Architecture Framework},
  institution = {3rd Generation Partnership Project (3GPP)},
  number      = {TS 28.533},
  year        = {2024},
  note        = {Release 17}
}

@techreport{ts28541,
  author      = {{3GPP}},
  title       = {Management and Orchestration; {5G} Network Resource Model ({NRM}); Stage 2 and Stage 3},
  institution = {3rd Generation Partnership Project (3GPP)},
  number      = {TS 28.541},
  year        = {2024},
  note        = {Release 17}
}

@techreport{ts29522,
  author      = {{3GPP}},
  title       = {{5G} System; Network Exposure Function Northbound {APIs}; Stage 3},
  institution = {3rd Generation Partnership Project (3GPP)},
  number      = {TS 29.522},
  year        = {2025},
  note        = {Release 17}
}

@techreport{oran_slicing_arch,
  author      = {{O-RAN Alliance}},
  title       = {{O-RAN} Slicing Architecture},
  institution = {O-RAN Alliance, Working Group 1},
  number      = {O-RAN.WG1.TS.Slicing-Architecture-R004-v14.01},
  year        = {2024}
}

@techreport{oran_slicing_study,
  author      = {{O-RAN Alliance}},
  title       = {Study on {O-RAN} Slicing},
  institution = {O-RAN Alliance, Working Group 1},
  number      = {O-RAN.WG1.Study-on-O-RAN-Slicing-v02.00},
  year        = {2022}
}

@techreport{oran_a1,
  author      = {{O-RAN Alliance}},
  title       = {{O-RAN} {A1} Interface: General Aspects and Principles},
  institution = {O-RAN Alliance, Working Group 2},
  number      = {O-RAN.WG2.A1GAP},
  year        = {2023}
}

@techreport{oran_e2ap,
  author      = {{O-RAN Alliance}},
  title       = {{O-RAN} {E2} Application Protocol ({E2AP})},
  institution = {O-RAN Alliance, Working Group 3},
  number      = {O-RAN.WG3.E2AP},
  year        = {2023}
}

@techreport{oran_r1,
  author      = {{O-RAN Alliance}},
  title       = {{O-RAN} {R1} Interface: General Aspects and Principles},
  institution = {O-RAN Alliance, Working Group 2},
  number      = {O-RAN.WG2.R1GAP},
  year        = {2023}
}

@techreport{oran_oam_arch,
  author      = {{O-RAN Alliance}},
  title       = {{O-RAN} Operations and Maintenance Architecture},
  institution = {O-RAN Alliance, Working Group 10},
  number      = {O-RAN.WG10.OAM-Architecture},
  year        = {2023}
}

@techreport{etsi_nfv_man001,
  author      = {{ETSI}},
  title       = {Network Functions Virtualisation ({NFV}); Management and Orchestration},
  institution = {European Telecommunications Standards Institute},
  number      = {ETSI GS NFV-MAN 001 V1.1.1},
  year        = {2014}
}

@techreport{etsi_nfv_ifa036,
  author      = {{ETSI}},
  title       = {Network Functions Virtualisation ({NFV}); Management and Orchestration; Report on {NFV-MANO} Architectural Framework Options},
  institution = {European Telecommunications Standards Institute},
  number      = {ETSI GS NFV-IFA 036},
  year        = {2021}
}

@techreport{etsi_nfv_ifa040,
  author      = {{ETSI}},
  title       = {Network Functions Virtualisation ({NFV}); Management and Orchestration; Requirements and Interfaces Specification for Containerized {VNF} Management},
  institution = {European Telecommunications Standards Institute},
  number      = {ETSI GS NFV-IFA 040},
  year        = {2021}
}

@techreport{etsi_gs_nfv_006,
  author      = {{ETSI}},
  title       = {Network Functions Virtualisation ({NFV}) Release 5; Management and Orchestration; Architectural Framework Specification},
  institution = {European Telecommunications Standards Institute},
  number      = {ETSI GS NFV 006 V5.2.1},
  year        = {2024}
}

@techreport{ngmn_cloud_native_manifesto,
  author      = {{NGMN Alliance}},
  title       = {Cloud Native Manifesto: An Operator View},
  institution = {Next Generation Mobile Networks Alliance},
  year        = {2023},
  note        = {Version 1.0, approved 6 September 2023}
}

@techreport{ITU-R-M2160,
  author      = {{ITU-R}},
  title       = {Framework and Overall Objectives of the Future Development of {IMT} for 2030 and Beyond},
  institution = {International Telecommunication Union, Radiocommunication Sector},
  number      = {Recommendation ITU-R M.2160-0},
  address     = {Geneva, Switzerland},
  year        = {2023}
}

@misc{oai,
  author       = {{OpenAirInterface Software Alliance}},
  title        = {{OpenAirInterface} Repository},
  year         = {2026},
  howpublished = {\url{https://gitlab.eurecom.fr/oai/openairinterface5g}},
  note         = {Accessed: Apr. 30, 2026}
}

@misc{open5gs,
  author       = {{Open5GS Project}},
  title        = {{Open5GS} Repository},
  year         = {2026},
  howpublished = {\url{https://github.com/open5gs/open5gs}},
  note         = {Accessed: Apr. 30, 2026}
}

@misc{operatorpattern,
  author       = {{Kubernetes Documentation}},
  title        = {Operator Pattern},
  year         = {2026},
  howpublished = {\url{https://kubernetes.io/docs/concepts/extend-kubernetes/operator/}},
  note         = {Accessed: Apr. 30, 2026}
}

@misc{kubernetes_controllers,
  author       = {{Kubernetes Documentation}},
  title        = {Controllers},
  year         = {2026},
  howpublished = {\url{https://kubernetes.io/docs/concepts/architecture/controller/}},
  note         = {Accessed: Jun. 3, 2026}
}

@misc{operatorsdk,
  author       = {{Operator Framework}},
  title        = {Operator {SDK} Repository},
  year         = {2026},
  howpublished = {\url{https://github.com/operator-framework/operator-sdk}},
  note         = {Accessed: Apr. 30, 2026}
}

@misc{olm,
  author       = {{Operator Framework}},
  title        = {Operator Lifecycle Manager},
  year         = {2024},
  howpublished = {\url{https://github.com/operator-framework/operator-lifecycle-manager}},
  note         = {Accessed: Apr. 23, 2026}
}

@misc{inotify,
  author       = {{Linux man-pages Project}},
  title        = {inotify(7): Monitoring Filesystem Events},
  year         = {2025},
  howpublished = {\url{https://man7.org/linux/man-pages/man7/inotify.7.html}},
  note         = {Accessed: Apr. 30, 2026}
}

@misc{tc,
  author       = {{Linux man-pages Project}},
  title        = {tc(8): Show / Manipulate Traffic Control Settings},
  year         = {2025},
  howpublished = {\url{https://man7.org/linux/man-pages/man8/tc.8.html}},
  note         = {Accessed: Apr. 30, 2026}
}

@misc{onap_so_architecture,
  author       = {{ONAP Project}},
  title        = {{ONAP} Service Orchestrator Architecture},
  year         = {2026},
  howpublished = {\url{https://docs.onap.org/projects/onap-so/en/latest/architecture/architecture.html}},
  note         = {Accessed: Jun. 3, 2026}
}

@misc{onap_so_github,
  author       = {{ONAP Project}},
  title        = {{ONAP} Service Orchestrator Repository},
  year         = {2026},
  howpublished = {\url{https://github.com/onap/so}},
  note         = {Accessed: Jun. 3, 2026}
}

@misc{azure_compensating_transaction,
  author       = {{Microsoft Azure Architecture Center}},
  title        = {Compensating Transaction Pattern},
  year         = {2026},
  howpublished = {\url{https://learn.microsoft.com/en-us/azure/architecture/patterns/compensating-transaction}},
  note         = {Accessed: Jun. 3, 2026}
}

@misc{osm,
  author       = {{ETSI OSM}},
  title        = {Open Source {MANO}},
  year         = {2026},
  howpublished = {\url{https://osm.etsi.org/}},
  note         = {Accessed: Jun. 3, 2026}
}

@misc{osm_source_code,
  author       = {{ETSI OSM}},
  title        = {{OSM} Lifecycle Manager},
  year         = {2026},
  howpublished = {\url{https://osm.etsi.org/gitlab/osm/lcm}},
  note         = {Accessed: Jun. 5, 2026}
}

@misc{tacker,
  author       = {{OpenStack}},
  title        = {Tacker: {OpenStack} {NFV} Orchestration},
  year         = {2026},
  howpublished = {\url{https://wiki.openstack.org/wiki/Tacker}},
  note         = {Accessed: Jun. 3, 2026}
}

@misc{tacker_source_code,
  author       = {{OpenStack}},
  title        = {Tacker Source Code},
  year         = {2026},
  howpublished = {\url{https://github.com/openstack/tacker}},
  note         = {Accessed: Jun. 5, 2026}
}

@misc{nephio,
  author       = {{Nephio Project}},
  title        = {Nephio: Cloud Native Network Automation},
  year         = {2026},
  howpublished = {\url{https://nephio.org/}},
  note         = {Accessed: Jun. 3, 2026}
}

@misc{nephio_source_code,
  author       = {{Nephio Project}},
  title        = {Nephio Source Code},
  year         = {2026},
  howpublished = {\url{https://github.com/nephio-project/nephio}},
  note         = {Accessed: Jun. 5, 2026}
}

@misc{chaosmesh,
  author       = {{Chaos Mesh Authors}},
  title        = {Chaos Mesh: A Powerful Chaos Engineering Platform for {Kubernetes}},
  year         = {2026},
  howpublished = {\url{https://chaos-mesh.org/}},
  note         = {Accessed: Jun. 18, 2026}
}

\end{document}